\documentclass[conference, 10pt]{IEEEtran}
\IEEEoverridecommandlockouts
\usepackage{cite}
\usepackage{amsmath,amssymb, amsfonts, bbold, dsfont, xfrac}
\usepackage{algorithm}
\usepackage{algorithmic}
\usepackage{graphicx}
\usepackage[font=footnotesize]{caption}
\usepackage[font=footnotesize]{subcaption}

\usepackage{textcomp}
\usepackage{hyperref}
\usepackage{xcolor}
\usepackage{comment}
\usepackage{anyfontsize}
\usepackage{booktabs}
\usepackage{multirow}

\usepackage{pgfplots}
\pgfplotsset{compat=1.16}
\pgfplotsset{yticklabel style={text width=1.2em,align=right}}
\definecolor{matblue}{rgb}{0, 0.4470, 0.7410}
\definecolor{matred}{rgb}{0.85, 0.3250, 0.0980}
\definecolor{matorange}{rgb}{0.9290, 0.6940, 0.1250}
\definecolor{matviolet}{rgb}{0.4940, 0.1840, 0.5560}
\definecolor{matgreen}{rgb}{0.4660, 0.6740, 0.1880}
\definecolor{matskyblue}{rgb}{0.3010, 0.7450, 0.9330}
\definecolor{matburgundy}{rgb}{0.6350, 0.0780, 0.1840}
\definecolor{matteal}{rgb}{0, .9, .9}

\newcommand{\vect}[1]{\boldsymbol{ #1 }}
\newcommand{\cvar}[2]{\text{CVaR}^{#2}\left[#1 \right]}

\newcommand\numberthis{\addtocounter{equation}{1}\tag{\theequation}}

\newtheorem{theorem}{Theorem}

\newtheorem{remark}{Remark}

\begin{document}

\title{Distributionally Robust Power Policies for\\Wireless Systems under Power Fluctuation Risk
\thanks{Contact: \{\href{mailto:gokberk.yaylali@yale.edu}{gokberk.yaylali}, \href{mailto:dionysis.kalogerias@yale.edu}{dionysis.kalogerias}\}@yale.edu.}
\thanks{This work is supported by the NSF under grant CCF 2242215.}
}
\author{\IEEEauthorblockN{Gokberk Yaylali and Dionysis Kalogerias}
\IEEEauthorblockA{\textit{Department of Electrical and Computer Engineering}\\\textit{Yale University}}
}

\maketitle

\begin{abstract}
Modern wireless communication systems necessitate the development of cost-effective resource allocation strategies, while ensuring maximal system performance.
While commonly realizable via efficient waterfilling schemes, ergodic-optimal policies 
often exhibit instantaneous resource constraint fluctuations as a result of fading variability, violating prescribed specifications possibly within unacceptable margins, inducing further operational challenges and/or costs. On the other extent, short-term-optimal policies --commonly based on deterministic waterfilling--, while strictly maintaining operational specifications, are not only impractical and computationally demanding, but also suboptimal in a long-term sense.
To address these challenges, we introduce a novel distributionally robust version of a classical point-to-point interference-free multi-terminal constrained stochastic resource allocation problem, by leveraging the Conditional Value-at-Risk (CVaR) as a coherent measure of \textit{power policy fluctuation risk}. We derive closed-form dual-parameterized expressions for the CVaR-optimal resource policy, along with corresponding optimal CVaR quantile levels by capitalizing on (sampling) the underlying fading distribution. We subsequently develop two dual-domain schemes ---one model-based and one model-free--- to iteratively determine a globally-optimal resource policy. Our numerical simulations confirm the remarkable effectiveness of the proposed approach, also revealing an almost-constant character of the CVaR-optimal policy and at rather minimal ergodic rate optimality loss.
\end{abstract}


\section{Introduction}
\label{sec:introduction}
We revisit the classical problem of optimal resource allocation in point-to-point multi-terminal communication networks operating over random fading channels 
with no cross-interference. Modern applications demand for maximal system performance and efficient utilization of resources, while also ensuring that certain \textit{operational} specifications are met. For instance, 
service providers are often concerned with spontaneous fluctuations in the allocated resources, e.g., power, which are costly due to over- or under-utilization of the underlying power grid and related provider-to-provider agreements. In fact, while ergodic-optimal (power) policies (typically implemented via waterfilling schemes in standard settings \cite{ribeiro_optimal_2012}) exhibit maximal performance in expectation and admit efficient implementations, they are prone to severe instantaneous fluctuations due to the statistical variability of channel fading, leading to operational challenges and/or increased costs. On the other extent, even though short-term optimal allocation schemes (such as deterministic waterfilling) strictly regulate operational specifications (i.e., constraints are met with certainty), they are not only suboptimal in a long-term sense, but also impractical: As an example, deterministic waterfilling suffers from heavy computational burden, since meeting instantaneous power constraints requires perpetual computation (i.e., solving one optimization problem per channel realization).

\textit{Distributionally Robust Optimization (DRO)} approaches to optimal resource allocation have been recently emerging \cite{li_efficient_2021, bennis_ultrareliable_2018, vu_ultra_reliable_2018, yang_localization_assisted_2024, batewela_risk_sensitive_2020, azari_risk-aware_2019, chai_u_reliable_v2x_2024, wang_jspa_2023, wu_rra_v2x_2023, cui_dro_mec_2023, wu_rra_vc_2021, ling_dro_offloading_2020, lie_dro_jammer_2019, gong_dro_relay_2016, li_dro_beamforming_2014, li_ra_dro_mec_2023}, including work by the authors \cite{yaylali_robust_2023, kalogerias_strongdual_2023, yaylali_stochastic_2024}, offering a new and fresh perspective on ultimately achieving explicit, tunable, and feasible operational robustness and (ultra-)reliability in wireless systems design. DRO involves objectives/constraints comprised of \textit{worst-case expectations} over \textit{an ambiguity set of distributions} \cite{rahimian_dro_2022, lin_dro_2022, shapiro_lectures_2021, kuhn_dro_2024}. In the context of wireless systems, optimal resource policies resulting from DRO models can provide maximally robust/reliable performance over an ambiguity set within the proximity of the (unknown) distribution of channel fading. From a complementary perspective, DRO may also be seen as a way to \textit{relax instantaneous objectives/constraints}, sometimes leading to favorable and analytically tractable problem formulations via convex functional duality \cite{ang_dual_2017}; we will discuss this aspect in more detail later on.

The notion of a \textit{risk measure} \cite{shapiro_lectures_2021, rockafellar_optimization_2000} is of key significance in the context of DRO. In fact, under very general conditions DRO functionals (e.g., used as problem objectives/constraints) arise as \textit{coherent risk measures} \cite[Chapters 6 and 7]{shapiro_lectures_2021}, since every such risk measure can be represented (in duality) as a supremum of expectations over a set of distributions \cite[Section 6.3]{shapiro_lectures_2021}, i.e., a coherent risk measure $\rho(\cdot)$ on an integrable random cost $\xi$ can be written as $\rho(\xi) = \sup_{Q \in \mathds{U}} \mathbb{E}_{Q}\left[ \xi \right]$, where $\mathds{U}$ is a corresponding set of probability distributions. Thus, DRO problems may also be seen as equivalent coherent \textit{risk-averse problems}, and vice versa (under general conditions).


\textit{In this paper}, we investigate a novel distributionally robust version of a classical resource allocation problem (as discussed earlier), formulated by relaxing desirable \textit{instantaneous} power constraints via suprema of expectations over \textit{fictitious} distributional balls of the form $\mathds{U}^\epsilon$ with radius $\epsilon$, centered at the true but unknown channel fading distribution. The selection of $\mathds{U}^\epsilon$ and $\epsilon$ regulates the degree of relaxation, i.e., $\epsilon \to 0$ (or when $\epsilon$ is sufficiently small) reduces to an expectation (ergodic setting), while $\epsilon \to \infty$ yields the essential supremum over the fading distribution, ensuring deterministic feasibility (initial instantaneous setting). The ball $\mathds{U}^\epsilon$ is tacitly selected herein by capitalizing on the dual representation of the Conditional Value-at-Risk (CVaR) \cite{rockafellar_optimization_2000, shapiro_lectures_2021}, the latter thus being adopted as a \textit{coherent measure of power policy fluctuation risk}. 

While DRO relaxations may in general result in a much more difficult resource allocation problems as compared with both the ergodic and deterministic extremes, we rigorously show that the consideration of CVaR (through its ambiguity set) and its favorable properties in fact enable the derivation of \textit{globally optimal} distributionally robust allocation \textit{policies in closed forms} (i.e., in the fashion of ergodic waterfilling). As a result, optimally minimal policy variability is ensured which is always feasible and tunable at will (by properly selecting the radius $\epsilon$), thus approximating the deterministic constraint setting with arbitrary precision. The proposed optimal policies depend on (\textit{deterministic}) CVaR-auxiliary and dual variables (Lagrange multipliers), which can be efficiently determined respectively via bisection and either dual subgradient descent or a basic primal-dual scheme ---both developed herein, reminiscent of and extending standard ergodic waterfilling. 

In summary, the proposed approach is able to achieve the implementation efficiency of ergodic-optimal resource allocation, while ensuring almost deterministic satisfaction of power constraints.
The efficacy of our approach is corroborated by a series of numerical simulations utilizing two common network utilities, also observing very minimal (ergodic) rate optimality loss, as a result of constraining power policy variability (via the DRO-CVaR relaxation).

\section{System Model}
\label{sec:system_model}

We consider a $N_U$-terminal parallel point-to-point communication channel with no cross-interference. Also, we assume perfect CSI for concreteness. While models including cross-interference refer to a larger variety of practical scenarios, the interference-free model studied herein is motivated mainly by its simplicity as well as analytical tractability, also facilitating the exposition of the novel aspect of our approach.
Cross-interference channels are certainly interesting, and they can be the subject of future work. We model the resource (here power) allocation policy as a vector $\vect{p}(\vect{h}) \succeq \vect{0}$, where $\vect{h}$ is the instantaneous fading vector, whose elements $h_i \in \mathcal{H}, i \in [N_U]$ correspond to the fading coefficients of parallel links, with cumulative distribution function (CDF) $F_{h_i}(\cdot)$. 
The instantaneous transmission rate of terminal $i$ of the network is modeled as
\begin{equation}
\label{eqn:rate_function}
r_i(p_i(\vect{h}), h_i) \triangleq \log\left( 1 + \frac{p_i(\vect{h}) \cdot h_i^2}{\sigma_i^2} \right),
\end{equation}
where $\sigma_i^2$ is the noise variance of the corresponding transmission link. In a very classical ergodic setting, the problem of optimal power allocation may be formulated as
\begin{equation}
\label{eqn:problem_formulation_ergodic_0}
\begin{aligned}
P_E^* = \underset{\vect{x} \in \mathcal{X},\ \vect{p}\succeq \vect{0}}{\mathrm{maximize}} \quad & f_0(\vect{x})\\
\mathrm{subject\ to} \quad & \vect{x} \preceq \mathbb{E}\left[ \vect{r}(\vect{p}(\vect{h}),\vect{h}) \right]\\
& 0 \leq P_0 - \sum_{i=1}^{N_U} \mathbb{E}[p_i(\vect{h})],
\end{aligned}
\end{equation}
where $f_0$ is a known and concave utility function, $\vect{x}$ is a vector measuring mean-ergodic system performance, $\mathcal{X}$ is a convex set, and $P_0$ is a total power budget. Problem \eqref{eqn:problem_formulation_ergodic} is well-studied in the literature \cite{ribeiro_ergodic_2010, ribeiro_optimal_2012, ribeiro_separation_2010, cover_elements_2005}, and a globally optimal solution is readily available via the well-known \textit{waterfilling algorithm} \cite{cover_elements_2005, ribeiro_optimal_2012}. However, although ergodic-optimal policies perform optimally in expectation, they are prone to fluctuations beyond $P_0$, e.g., power spikes, particularly over heavy-tailed fading channels. To mitigate the effects of dispersive channel fading, particularly for applications where power constraints are strictly imposed,
an alternative problem may formulated as
\begin{equation}
\label{eqn:problem_formulation_ergodic}
\begin{aligned}
P_I^* = \underset{\vect{x} \in \mathcal{X},\ \vect{p}\succeq \vect{0}}{\mathrm{maximize}} \quad & f_0(\vect{x})\\
\mathrm{subject\ to} \quad & \vect{x} \preceq \mathbb{E}\left[ \vect{r}(\vect{p}(\vect{h}),\vect{h}) \right]\\
& 0 \leq P_0 - \sum_{i=1}^{N_U} p_i(\vect{h}), \quad \mathrm{a.e.},
\end{aligned}
\end{equation}
ensuring \textit{instantaneous} total power feasibility and eliminating short-term power fluctuation. However, this power constraint is now essentially of infinite dimension, easily rendering \eqref{eqn:problem_formulation_ergodic} impractical and computationally demanding (see Section~\ref{sec:introduction}). 

To deal with those issues, we propose a \textit{DRO relaxation on the power constraint}, resulting in the problem
\begin{equation}
\label{eqn:problem_formulation_instant}
\begin{aligned}
P_R^*(\vect{\epsilon}) = \underset{\vect{x} \in \mathcal{X},\ \vect{p}\succeq \vect{0}}{\mathrm{maximize}} \quad & f_0(\vect{x})\\
\mathrm{subject\ to} \quad & \vect{x} \preceq \mathbb{E}\left[ \vect{r}(\vect{p}(\vect{h}),\vect{h}) \right]\\
& 0 \leq P_0 - \sum_{i=1}^{N_U} \sup_{ Q \in \mathds{U}_i^{\epsilon_i}}\ \mathbb{E}_{Q}[p_i(\vect{h})],
\end{aligned}
\end{equation}
over balls of \textit{testing} \textit{fictitious} distributions $\mathds{U}_i^{\epsilon_i}$ of radii $\epsilon_i$, each containing the true fading distribution as its ``center''; we will make this rigorous shortly. The reader might note that, at this point, whether problem \eqref{eqn:problem_formulation_instant} is actually advantageous over \eqref{eqn:problem_formulation_ergodic} is questionable; in general, \eqref{eqn:problem_formulation_instant} may be much harder to tackle than \eqref{eqn:problem_formulation_ergodic} (due to the suprema in the power constraint).

In this work, we exploit the CVaR to facilitate tackling the DRO \eqref{eqn:problem_formulation_ergodic}. Specifically, the CVaR is defined for an integrable random cost $\xi(\cdot)$ as \cite{shapiro_lectures_2021, rockafellar_optimization_2000}
\begin{equation}
\label{eqn:cvar_definition}
\cvar{\xi}{\varphi} \triangleq \inf_{z \in \mathbb{R}} \left\{ z + \frac{1}{\varphi} \mathbb{E}\left[ (\xi - z)_+ \right] \right\},
\end{equation}
 where $\varphi \in (0,1]$ is the corresponding \textit{confidence level}, and $(\cdot)_+ = \max\{\cdot, 0 \}$. The CVaR expresses the \textit{mean
of the worst \mbox{$\alpha\cdot100\%$} values of} \mbox{$\xi$}, and admits the perhaps more explanatory representation as a conditional expectation
\[
\cvar{\xi}{\varphi}=
\mathbb{E}[\xi|\xi \ge z_\varphi^*],\quad\text{where} ~ { P}(\xi \geq z_\varphi^*) = \varphi,
\]
whenever $\xi$ is continuously distributed (i.e., admits a density). In words, the CVaR measures the expected loss of $\xi$ in the upper $\varphi$-quantile of the underlying distribution, or conditioned to the upper tail of probability equal to $\varphi$.  The CVaR is a coherent risk measure ---i.e., a convex, monotone, translation equivariant, and positively homogeneous functional \cite[Chapter 6]{shapiro_lectures_2021}---, and generalizes the expectation since
\begin{equation*}
\lim_{\varphi \to 0}\ \cvar{\xi}{\varphi} 
\triangleq \cvar{\xi}{0} = \mathrm{ess} \sup \xi, \,\,\,\, \cvar{\xi}{1} = \mathbb{E}[\xi],
\end{equation*}
also being monotone increasing in $\varphi \in [0,1]$. Being a coherent risk measure, the CVaR admits the \textit{dual representation}
\begin{align*}
\cvar{\xi}{\varphi} &= \sup_{ Q \in \mathds{U}^{1/\varphi}}\ \mathbb{E}_{Q} \left[ \xi \right],
\end{align*}
where the ambiguity set $\mathds{U}^{1/\varphi}$ is takes the form
\begin{align*}
\mathds{U}^{1/\varphi} &= \left\{ Q \,\bigg| \,
\dfrac{\mathrm{d}Q}{\mathrm{d}P}
\in \left[ 0, \dfrac{1}{\varphi} \right]\ \text{a.e.-}P \right\}
\\
&= \left\{ Q \,\bigg| \,
\log \dfrac{\mathrm{d}Q}{\mathrm{d}P}
\le \log\dfrac{1}{\varphi}\ \text{a.e.-}P \right\},
\end{align*}
where $P$ denotes the true (reference) distribution. In words, $\mathds{U}^{1/\varphi}$ contains all distributions whose $\log$-likelihood ratio over $P$ does not exceed $\log(1/\varphi)$ in the essential supremum (Renyi) divergence. In that sense, $\mathds{U}^{1/\varphi}$ is a distributional ball centered at  $P$, with radius $\log(1/\varphi)$. Note that hereafter we slightly abuse notation by the identification $\mathds{U}^{1/\varphi}\equiv \mathds{U}^{\log(1/\varphi)}$.

By exploiting the CVaR through its dual representation in the DRO problem \eqref{eqn:problem_formulation_instant}, we obtain the \textit{risk-averse reformulation}
\begin{equation}
\label{eqn:problem_formulation_RA}
\begin{aligned}
P^*_R(\vect{\varphi}) = \underset{\vect{x} \in \mathcal{X},\ \vect{p}\succeq \vect{0}}{\mathrm{maximize}} \quad & f_0(\vect{x})\\
\mathrm{subject\ to} \quad & \vect{x} \preceq \mathbb{E}\left[ \vect{r}(\vect{p}(\vect{h}),\vect{h}) \right]\\
& 0 \leq P_0 - \sum_{i=1}^{N_U} \cvar{p_i(\vect{h})}{\varphi_i},
\end{aligned}
\end{equation}
where $\varphi_i\in (0,1]$ for all $i$ and $\vect{\varphi}$ is defined accordingly. Note that all CVaRs are taken relative to the distribution of $\vect{h}$.

At this point, we emphasize that one could start directly with the risk-averse problem \eqref{eqn:problem_formulation_RA} (see \cite{yaylali_robust_2023, yaylali_stochastic_2024}), which may be naturally interpreted as a \textit{tightening of the power constraint} in the ergodic problem \eqref{eqn:problem_formulation_ergodic_0}. 
Nevertheless, the DRO problem \eqref{eqn:problem_formulation_instant} is more general and probably more natural and intuitive to start with (i.e., relaxing the constraint of \eqref{eqn:problem_formulation_ergodic}), and the incorporation of CVaR may be seen as a special case (albeit rather useful, as discussed later). 
It easily follows that, for every $\vect{\varphi}\in [0,1]^{N_U}$,
\begin{align*}
P^*_I \le P^*_R(\vect{\varphi}) \le P^*_E,
\end{align*}
with equalities achieved for $\vect{\varphi} = \vect{0}$ and $\vect{\varphi} = \vect{1}$, respectively.

Substituting the CVaR formulation \eqref{eqn:cvar_definition}, we may re-express the DRO problem \eqref{eqn:problem_formulation_RA} as
\begin{align*}
P^*_R(\vect{\varphi}) \hspace{-1bp}=\hspace{-1bp} \underset{ \vect{x} \in \mathcal{X}, \vect{p}\succeq \vect{0}, \vect{z}}{\mathrm{maximize}} \quad & f_0(\vect{x})\\
\mathrm{subject\ to} \quad & \vect{x} \preceq \mathbb{E}\left[ \vect{r}(\vect{p}(\vect{h}),\vect{h}) \right] \numberthis \label{eqn:problem_formulation_final}\\
& 0 \leq P_0 \hspace{-1bp}-\hspace{-1bp} \sum_{i=1}^{N_U} z_i \hspace{-1bp}+\hspace{-1bp} \frac{1}{\varphi_i}\mathbb{E}\left[ (p_i(\vect{h}) \hspace{-1bp}-\hspace{-1bp} z_i)_+ \right] ,
\end{align*}
where the vector $\vect{z}$ collects CVaR auxiliary variables $z_i$. Problem \eqref{eqn:problem_formulation_final} is convex due to the structure of the CVaR. However, since fading coefficients attain values from a continuum $\mathcal{H}$, \eqref{eqn:problem_formulation_final} is infinite-dimensional, making it challenging to solve. 

Under the assumption of certain constraint qualifications, particularly Slater's condition (i.e., strict feasibility, \textit{assumed hereafter}), we observe strong duality (also implying no duality gap) for \eqref{eqn:problem_formulation_final}. This allows us to work with the dual problem of \eqref{eqn:problem_formulation_final} within the  framework of Lagrangian duality. To this end, the Lagrangian function associated with \eqref{eqn:problem_formulation_final} is defined as
\vspace{-2bp}
\begin{multline*}
\mathcal{L}(\vect{x}, \vect{p}, \vect{z}, \vect{\Lambda}) \triangleq f_0(\vect{x}) + \vect{\lambda}^\top \left( \mathbb{E}\left[ \vect{r}(\vect{p}(\vect{h},\vect{h}) \right] - \vect{x} \right)\\
+ \mu  P_0 - \sum_{i=1}^{N_U} \left( z_i + \frac{1}{\varphi_i}\mathbb{E}\left[ (p_i - z_i)_+ \right]  \right),
\end{multline*}
where $\vect{\Lambda} = (\vect{\lambda}, \mu) \succeq \vect{0}$ are Lagrange multipliers ---the dual variables corresponding to the constraints in \eqref{eqn:problem_formulation_final}. The dual function is defined as the maximization of the Lagrangian over the primal variables, i.e.,
\begin{align*}
\label{eqn:dual_function}
q(\vect{\Lambda}) \triangleq \sup_{\vect{x} \in \mathcal{X}, \vect{p}\succeq \vect{0}, \vect{z}}\ \mathcal{L}(\vect{x}, \vect{p}, \vect{z}, \vect{\Lambda}).
\end{align*}
We may then define the dual problem as
\begin{equation}
\label{eqn:dual_problem_formulation}
\begin{aligned}
D^*_R(\vect{\varphi}) &= \inf_{\vect{\Lambda} \succeq \vect{0}}\ q(\vect{\Lambda}),\\
&= \inf_{\vect{\Lambda} \succeq \vect{0}}\ \sup_{\vect{x} \in \mathcal{X}, \vect{p}\succeq \vect{0}, \vect{z}}\ \mathcal{L}(\vect{x}, \vect{p}, \vect{z}, \vect{\Lambda}).
\end{aligned}
\end{equation}
Although the primal DRO/risk-averse CVaR problem \eqref{eqn:problem_formulation_final} is infinite-dimensional, its dual \eqref{eqn:dual_problem_formulation} is finite-dimensional (in the dual variables), provided the availability of (implicit) optimal primal variables.
Next, we propose a distributionally robust resource allocation scheme to efficiently solve the minimax problem \eqref{eqn:dual_problem_formulation} to global optimality, in particular enjoying explicit closed forms as far as the resource policy $\vect{p}$ is concerned.

\section{Distributionally Robust Resource Allocation}
\label{sec:channel_invariant_RA}
We observe that the dual function is separable relative to $i \in [N_U]$ over the primal variables $(\vect{p},\vect{z})$, the latter also being separable from $\vect{x}$ (given respective dual variables). Thus, we can tackle the subproblems over $(p_i, z_i)$ and $\vect{x}$ separately.

\subsection{Optimal Ergodic Rate Vector}
\label{subsec:rate_vector_x}

The subproblem over the rate vector $\vect{x}$ is expressed as
\begin{equation*}
\sup_{\vect{x} \in \mathcal{X}} \left\{ f_0(\vect{x}) - \vect{\lambda}^\top \vect{x} \right\}.
\end{equation*}
The optimal solution set inherently depends on the concave utility $f_0$ and the dual variable $\vect{\lambda}$ 
and may be expressed as
\begin{equation}
\label{eqn:x_solution}
\vect{x}^*(\vect{\lambda}) \in \arg\max_{\vect{x} \in \mathcal{X}} \left\{ f_0(\vect{x}) - \vect{\lambda}^\top \vect{x} \right\}.
\end{equation}
We assume the existence and availability of such an optimal solution as a function of $\vect{\lambda}$, provided $f_0$ is analytically tractable and known. Several common utility functions allow us to obtain closed-form solutions and/or variable eliminations. For an example, the sumrate utility $f_0(\vect{x}) = \vect{w}^\top \vect{x}$ for a weight vector $\vect{w} \in \mathbb{R}_{++}^{N_U}$ yields an optimal dual solution $\vect{\lambda}^* \triangleq \vect{w}$, also eliminating the derivation of $\vect{x}$. Another choice involves the proportional fairness utility, i.e., $f_0(\vect{x}) = \vect{1}^\top \log(\vect{x})$, which yields an optimal solution $\vect{x}^* \triangleq \frac{1}{\vect{\lambda}}$; here element-wise logarithm and division notations are overloaded. Numerical simulations using these utilities are presented in Section~\ref{sec:performance_eval}.

\subsection{CVaR-Optimal Resource Policy}
\label{subsec:policy_p}

Leveraging the interchangeability principle for expectation (integration) \cite[Theorem~9.108]{shapiro_lectures_2021}, we consider the particular subproblem over policy $p_i$ for terminal $i \in [N_U]$, i.e.,
\begin{equation}
\label{eqn:p_subproblem}
\sup_{p_i \geq 0} \left\{ \lambda_i \log\left( 1 + \frac{p_i \cdot h_i^2}{\sigma_i^2}\right) - \frac{\mu}{\varphi_i}(p_i - z_i)_+ \right\}.
\end{equation}
We have the following central result.
\begin{theorem}[CVaR-Optimal Resource Policy]
A parameterized optimal solution to the resource policy subproblem \eqref{eqn:p_subproblem} for each terminal $i \in [N_U]$ may be expressed as
\begin{equation}
\label{eqn:pd_policy_soln}
p_i^*(h_i)= \max \left\{ \left( \frac{\lambda_i \varphi_i}{\mu} - \frac{\sigma_i^2}{h_i^2} \right)_+,\ (z_i)_+ \right\},
\end{equation}
whenever $\lambda_i$ and $\mu$ are not simultaneously zero, otherwise selecting $p_i^* = 0$ is optimal.
\end{theorem}
\begin{IEEEproof}
Denote the objective of \eqref{eqn:p_subproblem} as a function $F(p_i,\cdot)$. Notice that $F(p_i,\cdot)$ is concave in $p_i$, and becomes null for $\lambda_i$ and $\mu$ being simultaneously zero. For $\lambda_i = 0$ and $\mu > 0$, the subproblem stands trivial with an optimal policy selection of $p_i^* = (z_i)_+$. For $\lambda_i > 0$ and $\mu = 0$, the subproblem becomes unbounded and reduces to
\begin{equation*}
\sup_{p_i \geq 0} \left\{ \lambda_i \log\left( 1 + \frac{p_i \cdot h_i^2}{\sigma_i^2}\right) \right\},
\end{equation*}
with an ``infeasible solution'' of $p_i^* = \infty$. For $(\lambda_i, \mu) \succ \vect{0}$ --assumed hereafter, the subproblem evolves into
\begin{equation}
\label{eqn:p_subproblem_evolved}
\sup_{p_i}\ \lambda_i \log\left( 1 + \frac{p_i \cdot h_i^2}{\sigma_i^2}\right) - \frac{\mu}{\varphi_i}(p_i - z_i)_+ - \mathds{I}_{\mathbb{R}_+}(p_i),
\end{equation}
including the constraint on $p_i$, where
\begin{equation*}
\mathds{I}_{\mathbb{R}_+}(p_i) = \begin{cases}
0, & p_i \geq 0,\\
\infty, & p_i < 0.
\end{cases}
\end{equation*}
The subdifferential of \eqref{eqn:p_subproblem_evolved} may be expressed as $\partial_{p_i} F(p_i,\cdot) - \mathcal{N}_{\mathbb{R}_+}(p_i)$ where $\mathcal{N}_{\mathbb{R}_+}(p_i)$ is the normal cone of the nonnegative orthant $\mathbb{R}_+$, expressed as 
$\mathcal{N}_{\mathbb{R}_+}(p_i) = \{ n | n \in \mathbb{R}_- \ \mathrm{if}\ p_i = 0, n = 0 \ \mathrm{if}\ p_i > 0 \}$.
Then, every subgradient $g_{p_i}(p_i, \cdot)$ of the problem of \eqref{eqn:p_subproblem_evolved} may be expressed as
\begin{equation}
\label{eqn:p_subgradient}
g_{p_i}(p_i,\cdot) = \frac{\lambda_i h_i^2}{\sigma_i^2 + p_i \cdot h_i^2} - \frac{\mu}{\varphi_i} H[p_i - z_i] - n,
\end{equation}
where $n \in \mathcal{N}_{\mathbb{R}_+}(p_i)$, and $H[\cdot]$ is the Heaviside step multifunction. 
A zero-subgradient needs to be in the subdifferential as per the optimality condition, i.e., $0 \in \partial F(p_i^*, \cdot) - \mathcal{N}_{\mathbb{R}_+}(p_i^*)$. Assuming $z_i \leq 0$, $n \in \mathbb{R}_-$ is satisfied only for $p_i^* = 0, H[p_i^*] = C \in [0, 1]$, providing $\frac{\lambda_i \varphi_i h_i^2}{\mu \sigma_i^2} \leq C \leq 1$ and $\frac{\lambda_i \varphi_i}{\mu} - \frac{ \sigma_i^2}{h_i^2} \leq 0$ from \eqref{eqn:p_subgradient}. Otherwise, a zero-subgradient proves $p_i^* = \frac{\lambda_i \varphi_i}{\mu} - \frac{ \sigma_i^2}{h_i^2} > 0$ for $n = 0$, yielding altogether
\begin{equation}
\label{eqn:p_n_equation}
p_i^* = \left( \frac{\lambda_i \varphi_i}{\mu} - \frac{ \sigma_i^2}{h_i^2} \right)_+, \quad z_i \leq 0.
\end{equation}
Conversely, --assuming $z_i > 0$, a zero-subgradient satisfies
\begin{equation}
\label{eqn:p_zero_subgradient}
g_{p_i}(p_i^*, \cdot) = \frac{\lambda_i h_i^2}{\sigma_i^2 + p_i^* \cdot h_i^2} - \frac{\mu}{\varphi_i} H[p_i^* - z_i] = 0,
\end{equation}
Here, we observe that there are two scenarios for a subgradient to attain zero. For the first scenario, suppose a $p_i^* > 0$ exists such that $p_i^* > z_i \Longleftrightarrow H[p_i^* - z_i] = 1$. From \eqref{eqn:p_zero_subgradient}, we obtain
\begin{equation*}
p_i^* = \left( \frac{\lambda_i \varphi_i}{\mu} - \frac{\sigma_i^2}{h_i^2} \right)_+ > z_i,
\end{equation*}
providing a \textit{branch condition}. For the second scenario, suppose $p_i^* > 0$ exists such that $p_i^* = z_i \Longleftrightarrow H[p_i^* - z_i] = C \in [0,1]$. Then, from \eqref{eqn:p_zero_subgradient}, we arrive at
\begin{equation*}
g_{p_i}(p_i^*,\cdot) = \frac{\lambda_i h_i^2}{\sigma_i^2 + p_i^* \cdot h_i^2} - \frac{\mu}{\varphi_i}C = 0,
\end{equation*}
satisfying $p_i^* \geq \frac{\lambda_i \varphi_i}{\mu} - \frac{\sigma_i^2}{h_i^2}$, and providing the complementary branch condition. Note that  parameter $C$ may be expressed as
\begin{equation}
\label{eqn:p_C_expression}
C = H[p_i^* - z_i] = \frac{\lambda_i \varphi_i}{\mu}\frac{h_i^2}{\sigma_i^2 + p_i^* \cdot h_i^2},
\end{equation}
satisfying all scenarios. Combining both scenarios along with \eqref{eqn:p_n_equation} ultimately concludes the proof.
\end{IEEEproof}


\subsection{Optimal Value-at-Risk Levels}

Recalling the dual function, the subproblem related to $z_i$ for terminal $i \in [N_U]$ may be expressed as
\begin{equation}
\label{eqn:z_subproblem}
\sup_{z_i \in \mathbb{R}}\left\{ -\mu z_i + \mathbb{E}\left[ \lambda_i r_i(p_i^*, h_i) - \frac{\mu}{\varphi_i} (p_i^* - z_i)_+ \right] \right\},
\end{equation}
given the corresponding \textit{optimal} power policy. We present two dual-domain schemes to acquire the optimal $z_i^*$. First, a (quasi-)closed-form solution is derived, provided that (sampling) the underlying fading distribution is readily available. For a purely data-driven method, we also propose a primal-dual scheme utilizing subgradient ascent-based updates.

\subsubsection{Dual-Parameterized Optimal Value-at-Risk}
In case the fading distribution is available or can be sampled, the subproblem in \eqref{eqn:z_subproblem} yields a quasi-closed-form solution, as follows.
\begin{theorem}[Optimal Value-at-Risk]
\label{theorem:z_closed_form}
The optimal solution of subproblem \eqref{eqn:z_subproblem} for terminal $i \in [N_U]$ satisfies the inclusions
\begin{equation*}
z_i^* \hspace{-2pt}\in\hspace{-2pt} \begin{cases}
\{ 0 \}, & 1-\varphi_i \leq \mathbb{E}\left[ (1 - \kappa_i h_i^2) \cdot \mathds{1}_{\Bar{\mathcal{H}}(0)}(h_i) \right] \hspace{-1.5pt},\\
(0, \sigma_i^2 \kappa_i], & 1-\varphi_i \leq \mathbb{E}\left[ (1-C_{\sigma_i^2 \kappa_i}) \cdot \mathds{1}_{\Bar{\mathcal{H}}(\sigma_i^2 \kappa_i)}(h_i) \right] \hspace{-1.5pt},\\
(\sigma_i^2 \kappa_i, \infty), & 1-\varphi_i > \mathbb{E}\left[ (1-C_{\sigma_i^2 \kappa_i}) \cdot \mathds{1}_{\Bar{\mathcal{H}}(\sigma_i^2 \kappa_i)}(h_i) \right] \hspace{-1.5pt},
\end{cases}
\end{equation*}
where $\Bar{H}(z_i) = \sqrt{\frac{\sigma_i^2}{\sigma_i^2 \kappa_i - z_i}}$, the set $\Bar{\mathcal{H}}(z_i) = \left[0, \Bar{H}(z_i) \right]$, $\kappa_i = \frac{\lambda_i \varphi_i}{\mu \sigma_i^2}$, and $C_{z_i} = \frac{\sigma_i^2 \kappa_i h_i^2}{\sigma_i^2 + z_i h_i^2}$. The optimal solution $z_i^*$ satisfies
\begin{align*}
\mathbb{E}\left[ C_{z_i^*} \cdot \mathds{1}_{\Bar{\mathcal{H}}(z_i^*)} \right] + 1 - F_{h_i}\left( \Bar{H}(z_i^*) \right) &= \varphi_i, \quad z_i^* \in (0, \sigma_i^2 \kappa_i]\\
\mathbb{E}\left[ C_{z_i^*} \right] &= \varphi_i, \quad z_i^* \in [\sigma_i^2 \kappa_i,\infty).
\end{align*}
\end{theorem}

\begin{algorithm}[tbp]
\centering
\begin{algorithmic}
\STATE Choose initial values $\vect{x}^{(0)}, \vect{p}^{(0)}, \vect{z}^{(0)}, \vect{\Lambda}^{(0)}$.
\FOR{$n = 1$ \textbf{to} Process End}
\STATE Observe $\vect{h}^{(n)}$.
\STATE \textit{\# Primal Variables}
\STATE $\vect{\to}$ Set $z_i^*( \vect{\Lambda}^{(n-1)} )$ using Theorem~\ref{theorem:z_closed_form}, for all $i$.
\STATE $\vect{\to}$ Set $p_i^*( \vect{h}^{(n)}, \vect{\Lambda}^{(n-1)} )$ using \eqref{eqn:pd_policy_soln}, for all $i$.
\STATE $\vect{\to}$ Set $\vect{x}^*( \vect{\Lambda}^{(n-1)} )$ using \eqref{eqn:x_solution}.
\STATE \textit{\# Dual Variables}
\STATE $\vect{\to}$ Update $\vect{\Lambda}^{(n)}$ using \eqref{eqn:Lambda_dual_update} and \eqref{eqn:lambda_subgradient}.
\ENDFOR
\end{algorithmic}
\caption{Dual CVaR-Optimal Resource Allocation}
\label{alg:algorithm_1}
\end{algorithm}

\begin{IEEEproof}
Consider a function defined as
\begin{equation*}
F(p_i,z_i) \triangleq -\mu z_i + \lambda_i r_i(p_i, h_i) - \frac{\mu}{\varphi_i}(p_i - z_i)_+.
\end{equation*}
Note that $F(p_i,z_i)$ is \textit{jointly concave}, and for its subdifferential we may write
\begin{align*}
\partial_{z_i} \sup_{p_i \geq 0} F(p_i, z_i) &= \partial_{z_i} F(p_i, z_i) |_{p_i = p_i^*(\cdot)},\\
\partial_{z_i} \mathbb{E}\left[ \sup_{p_i \geq 0} F(p_i, z_i) \right] &= \mathbb{E}\left[ \partial_{z_i} F(p_i, z_i) |_{p_i = p_i^*(\cdot)} \right],\\
&= -\mu +\frac{\mu}{\varphi_i}\mathbb{E}\left[H[p_i^* - z_i] \right].
\end{align*}
Then, we may express a subgradient $g_{z_i} \in \partial_{z_i} \mathbb{E}\left[ F^*(z_i) \right]$ as
\begin{equation*}
g_{z_i} = -\mu + \frac{\mu}{\varphi_i}\mathbb{E}\left[ \mathds{1}_{ \{ p_i^* > z_i \} }(h_i) \right] + \frac{\mu}{\varphi_i}\mathbb{E}\left[ C_{z_i} \mathds{1}_{ \{ p_i^* = z_i \} }(h_i) \right],
\end{equation*}
where $C_{z_i} \in [0,1]$ is constrained by the selection of policy $p_i^*$ as in \eqref{eqn:p_C_expression}. Let $\Bar{H}(z_i) = \sqrt{\frac{\sigma_i^2}{\sigma_i^2 \kappa_i - z_i}}$, the set $\Bar{\mathcal{H}}(z_i) = \left[0, \Bar{H}(z_i) \right]$, 
and $\kappa_i = \frac{\lambda_i \varphi_i}{\mu \sigma_i^2}$ for simplicity. Then we may express the subgradient $g_{z_i}$ for different values of $z_i$, i.e.,
\vspace{-2pt}
\begin{multline*}
g_{z_i} = \frac{\mu(1-\varphi_i)}{\varphi_i} +\\
\begin{cases}
0, & z_i \in (-\infty,0),\\
\frac{\mu}{\varphi_i}\left(\mathbb{E}\left[ C_{0} \cdot \mathds{1}_{\Bar{\mathcal{H}}(0)} \right] - F_{h_i}\left( \Bar{H}(0) \right) \right), & z_i = 0,\\
\frac{\mu}{\varphi_i}\left(\mathbb{E}\left[ C_{z_i} \cdot \mathds{1}_{\Bar{\mathcal{H}}(z_i)} \right] - F_{h_i}\left( \Bar{H}(z_i) \right) \right), & z_i \in (0, \sigma_i^2 \kappa_i),\\
\frac{\mu}{\varphi_i}\left(\mathbb{E}\left[ C_{z_i} \right] - 1 \right), & z_i \in [\sigma_i^2 \kappa_i, \infty).
\end{cases}
\end{multline*}
Note that since $n \in \mathcal{N}_{\mathbb{R}_+}(0)$ can take any value in $\mathbb{R}_-$, $H[-z_i]$ lies in $\kappa_i h_i^2 \leq H[-z_i] \leq 1$ at $z_i=0$, allowing the corresponding $C$ parameter to be selected freely within the range. Here, $C = C_0 = \kappa_i h_i^2$ is picked as per the lower semi-continuity of $g_{z_i}$ for $z_i \geq 0$. Also note the subgradient $g_{z_i}$ is monotone decreasing and can attain zero --except for $z_i < 0$, showing an optimal $z_i$ is nonnegative. 

To obtain a zero subgradient, one first obtains the set in which the optimal $z_i$ lies. First, the optimal solution $z_i^* = 0$ provided that $1 - \varphi_i \leq \mathbb{E}\left[ (1 - \kappa_i h_i^2) \cdot \mathds{1}_{\Bar{\mathcal{H}}(0)}(h_i) \right]$ is satisfied. Further, $z_i^* \in (0, \sigma_i^2 \kappa_i]$ provided $1-\varphi_i \leq \mathbb{E}\left[ (1-C_{\sigma_i^2 \kappa_i}) \cdot \mathds{1}_{\Bar{\mathcal{H}}(\sigma_i^2 \kappa_i)}(h_i) \right]$ is satisfied. Otherwise, the optimal solution $z_i^*$ lies in $z_i^* \in (\sigma_i^2 \kappa_i, \infty)$. Combining the scenarios and the branches in $g_{z_i}$ concludes the proof.
\end{IEEEproof}

\begin{algorithm}[tbp]
\centering
\begin{algorithmic}
\STATE Choose initial values $\vect{x}^{(0)}, \vect{p}^{(0)}, \vect{z}^{(0)}, \vect{\Lambda}^{(0)}$.
\FOR{$n = 1$ \textbf{to} Process End}
\STATE Observe $\vect{h}^{(n)}$.
\STATE \textit{\# Primal Variables}
\STATE $\vect{\to}$ Update $z_i^{(n)}( \vect{\Lambda}^{(n-1)} )$ using \eqref{eqn:z_supergradient_ascent} and \eqref{eqn:z_stoch_subgradient}, for all $i$.
\STATE $\vect{\to}$ Set $p_i^*( \vect{h}^{(n)}, \vect{\Lambda}^{(n-1)} )$ using \eqref{eqn:pd_policy_soln}, for all $i$.
\STATE $\vect{\to}$ Set $\vect{x}^*( \vect{\Lambda}^{(n-1)} )$ using \eqref{eqn:x_solution}.
\STATE \textit{\# Dual Variables}
\STATE $\vect{\to}$ Update $\vect{\Lambda}^{(n)}$ using \eqref{eqn:Lambda_dual_update} and \eqref{eqn:lambda_subgradient}.
\ENDFOR
\end{algorithmic}
\caption{Primal-Dual CVaR-Optimal Resource Allocation}
\label{alg:algorithm_2}
\end{algorithm}

The solution $z_i^*$ may be obtained within arbitrary precision by a numerical method such as the bisection method, provided the fading distribution is readily available or can be sampled.

\subsubsection{Subgradient Update for Value-at-Risk}

In case the fading distribution is not  easily accessible, a purely data-driven update method for variables $z_i$ may be constructed by capitalizing on the stochastic supergradients on the objective in \eqref{eqn:z_subproblem}. 
A stochastic supergradient ascent update rule for $z_i,\ i \in [N_U]$ maximizing the objective of \eqref{eqn:z_subproblem} may be expressed as
\begin{equation}
\label{eqn:z_supergradient_ascent}
z_i^{(n)} \triangleq z_i^{(n-1)} + \varepsilon_{\vect{z}} \Tilde{g}_{z_i}^{(n-1)} \big(h_i^{(n)}, z_i^{(n-1)} \big), \quad n \geq 1,
\end{equation}
with a step size $\varepsilon_{\vect{t}}$ and an initialization $t_i^{(0)}$, where the stochastic subgradient $\Tilde{g}_{z_i}$ is expressed as
\begin{equation}
\label{eqn:z_stoch_subgradient}
\Tilde{g}_{z_i}(\cdot) = -\mu + \frac{\mu}{\varphi_i}C_{z_i}, \quad C_{z_i} \in [0, 1],
\end{equation}
where $C_{z_i}$ is constrained by the selection of policy $p_i^*$ in \eqref{eqn:p_C_expression},  provided that an iteration index $n \in \mathbb{N}$ and the processes $\big\{ h_i^{(n)} \big\}$ and $\big\{ z_i^{(n)} \big\}$ --$\big\{ \lambda_i^{(n)} \big\}$ and $\big\{ \mu^{(n)} \big\}$ are also implicitly assumed-- are readily available.

\begin{figure*}[t]
\begin{subfigure}[t]{.5\linewidth}
\centering
\begin{subfigure}[ht]{\linewidth}
\begin{tikzpicture}[trim axis right,baseline]
\begin{axis}[
width=\linewidth,
height=.4\linewidth,
ytick={0, 5, 10, 15, 20},
ylabel={Power},
xmin=0, xmax=100, ymin=-4, ymax=22, xmajorticks=false,
legend style={at={(1,0)},anchor=south east,
nodes={scale=0.4, transform shape}, legend columns = 4},
grid]
\addplot[matblue, very thick, mark=no] table[x=time,y=p1] {Data-n3/pf_n3_P15_phi0.90_0.85_0.80_var1.0_2.0_3.0_p_instances.txt};
\addplot[matred, very thick, mark=no] table[x=time,y=p2] {Data-n3/pf_n3_P15_phi0.90_0.85_0.80_var1.0_2.0_3.0_p_instances.txt};
\addplot[matorange, very thick, mark=no] table[x=time,y=p3] {Data-n3/pf_n3_P15_phi0.90_0.85_0.80_var1.0_2.0_3.0_p_instances.txt};
\addplot[matviolet, very thick, mark=no] table[x=time,y=sum_p] {Data-n3/pf_n3_P15_phi0.90_0.85_0.80_var1.0_2.0_3.0_p_instances.txt};
\legend{$\varphi = 0.90/\sigma^2 = 1.0$, $\varphi = 0.85/\sigma^2 = 2.0$, $\varphi = 0.80/\sigma^2 = 3.0$, Total }
\end{axis}
\end{tikzpicture}
\end{subfigure}
\begin{subfigure}[ht]{\linewidth}
\begin{tikzpicture}[trim axis right,baseline]
\begin{axis}[
width=\linewidth,
height=.4\linewidth,
ytick={0, 5, 10, 15, 20},
ylabel={Power},
xlabel={Time (Fading Realizations)},
xmin=0, xmax=100, ymin=-5, ymax=22, xmajorticks=true,
legend style={at={(1,0)},anchor=south east,
nodes={scale=0.4, transform shape}, legend columns = 4}, grid]
\addplot[matblue, very thick, mark=no] table[x=time,y=p1] {Data-n3/pf_n3_P15_phi1.00_1.00_1.00_var1.0_2.0_3.0_p_instances.txt};
\addplot[matred, very thick, mark=no] table[x=time,y=p2] {Data-n3/pf_n3_P15_phi1.00_1.00_1.00_var1.0_2.0_3.0_p_instances.txt};
\addplot[matorange, very thick, mark=no] table[x=time,y=p3] {Data-n3/pf_n3_P15_phi1.00_1.00_1.00_var1.0_2.0_3.0_p_instances.txt};
\addplot[matviolet, very thick, mark=no] table[x=time,y=sum_p] {Data-n3/pf_n3_P15_phi1.00_1.00_1.00_var1.0_2.0_3.0_p_instances.txt};
\legend{$\varphi = 1.00/\sigma^2 = 1.0$, $\varphi = 1.00/\sigma^2 = 2.0$, $\varphi = 1.00/\sigma^2 = 3.0$, Total }
\end{axis}
\end{tikzpicture}
\end{subfigure}
\label{fig:p_vals_pf}
\end{subfigure}
\begin{subfigure}[t]{.5\linewidth}
\centering
\begin{subfigure}[ht]{\linewidth}
\begin{tikzpicture}[trim axis right,baseline]
\begin{axis}[
width=\linewidth,
height=.4\linewidth,
ytick={0, 5, 10, 15, 20},
ylabel={Power},
xmin=0, xmax=100, ymin=-4, ymax=22, xmajorticks=false,
legend style={at={(1,0)},anchor=south east,
nodes={scale=0.4, transform shape}, legend columns = 4},
grid]
\addplot[matblue, very thick, mark=no] table[x=time,y=p1] {Data-n3/sr_n3_P15_phi0.90_0.85_0.80_var1.0_2.0_3.0_p_instances.txt};
\addplot[matred, very thick, mark=no] table[x=time,y=p2] {Data-n3/sr_n3_P15_phi0.90_0.85_0.80_var1.0_2.0_3.0_p_instances.txt};
\addplot[matorange, very thick, mark=no] table[x=time,y=p3] {Data-n3/sr_n3_P15_phi0.90_0.85_0.80_var1.0_2.0_3.0_p_instances.txt};
\addplot[matviolet, very thick, mark=no] table[x=time,y=sum_p] {Data-n3/sr_n3_P15_phi0.90_0.85_0.80_var1.0_2.0_3.0_p_instances.txt};
\legend{$\varphi = 0.90/\sigma^2 = 1.0$, $\varphi = 0.85/\sigma^2 = 2.0$, $\varphi = 0.80/\sigma^2 = 3.0$, Total }
\end{axis}
\end{tikzpicture}
\end{subfigure}
\begin{subfigure}[ht]{\linewidth}
\begin{tikzpicture}[trim axis right,baseline]
\begin{axis}[
width=\linewidth,
height=.4\linewidth,
ytick={0, 5, 10, 15, 20},
ylabel={Power},
xlabel={Time (Fading Realizations)},
xmin=0, xmax=100, ymin=-4, ymax=22, xmajorticks=true,
legend style={at={(1,0)},anchor=south east,
nodes={scale=0.4, transform shape}, legend columns = 4}, grid]
\addplot[matblue, very thick, mark=no] table[x=time,y=p1] {Data-n3/sr_n3_P15_phi1.00_1.00_1.00_var1.0_2.0_3.0_p_instances.txt};
\addplot[matred, very thick, mark=no] table[x=time,y=p2] {Data-n3/sr_n3_P15_phi1.00_1.00_1.00_var1.0_2.0_3.0_p_instances.txt};
\addplot[matorange, very thick, mark=no] table[x=time,y=p3] {Data-n3/sr_n3_P15_phi1.00_1.00_1.00_var1.0_2.0_3.0_p_instances.txt};
\addplot[matviolet, very thick, mark=no] table[x=time,y=sum_p] {Data-n3/sr_n3_P15_phi1.00_1.00_1.00_var1.0_2.0_3.0_p_instances.txt};
\legend{$\varphi = 1.00/\sigma^2 = 1.0$, $\varphi = 1.00/\sigma^2 = 2.0$, $\varphi = 1.00/\sigma^2 = 3.0$, Total }
\end{axis}
\end{tikzpicture}
\end{subfigure}
\label{fig:p_vals_sr}
\end{subfigure}
\caption{Allocated power for $3$-terminal network with proportional fairness (left) and  sumrate (right) utilities, distributionally robust (top) and ergodic (bottom).}
\label{fig:p_vals}
\end{figure*}
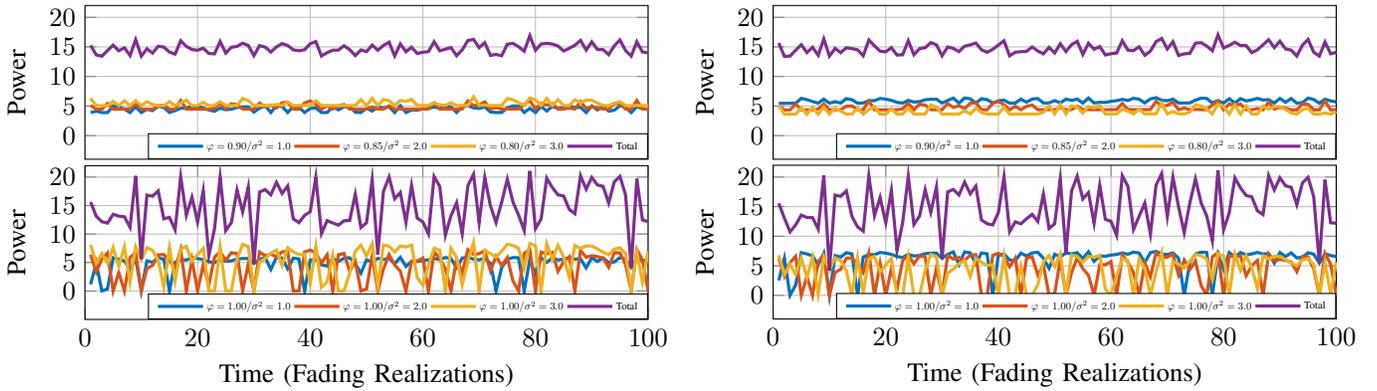
\begin{figure*}[t]
\begin{subfigure}[ht]{.5\linewidth}
\centering
\begin{subfigure}[ht]{\linewidth}
\begin{tikzpicture}[trim axis right,baseline]
\begin{axis}[
width=\linewidth,
height=.4\linewidth,
ylabel={CDF}, 
xmin=0, xmax=5, xmajorticks=false,
ymin=0, ymax=1.1, ytick = {0.5, 1},
legend image post style={scale=0.4},
legend style={at={(0.995,0.01)},anchor= south east,
nodes={scale=0.4, transform shape}}, grid]
\addplot[matblue, very thick] table[x=rate,y=cdf_rate1] {Data-n3/pf_n3_P15_phi0.90_0.85_0.80_var1.0_2.0_3.0_rate_CDF.txt};
\addplot[matred, very thick] table[x=rate,y=cdf_rate2] {Data-n3/pf_n3_P15_phi0.90_0.85_0.80_var1.0_2.0_3.0_rate_CDF.txt};
\addplot[matorange, very thick] table[x=rate,y=cdf_rate3] {Data-n3/pf_n3_P15_phi0.90_0.85_0.80_var1.0_2.0_3.0_rate_CDF.txt};
\addplot[matblue, very thick, dashed] table[x=rate,y=cdf_rate1] {Data-n3/pf_n3_P15_phi1.00_1.00_1.00_var1.0_2.0_3.0_rate_CDF.txt};
\addplot[matred, very thick, dashed] table[x=rate,y=cdf_rate2] {Data-n3/pf_n3_P15_phi1.00_1.00_1.00_var1.0_2.0_3.0_rate_CDF.txt};
\addplot[matorange, very thick, dashed] table[x=rate,y=cdf_rate3] {Data-n3/pf_n3_P15_phi1.00_1.00_1.00_var1.0_2.0_3.0_rate_CDF.txt};
\legend{$\varphi = 0.90/\sigma^2 = 1.0$, $\varphi = 0.85/\sigma^2 = 2.0$, $\varphi = 0.80/\sigma^2 = 3.0$, $\varphi = 1.00/\sigma^2 = 1.0$, $\varphi = 1.00/\sigma^2 = 2.0$, $\varphi = 1.00/\sigma^2 = 3.0$}
\end{axis}
\end{tikzpicture}
\end{subfigure}
\\
\begin{subfigure}[ht]{\linewidth}
\begin{tikzpicture}[trim axis right,baseline]
\begin{axis}[
width=\linewidth,
height=.4\linewidth,
ylabel={CDF}, xlabel={Instantaneous Rate},
xmin=0, xmax=5, xmajorticks=true,
ymin=0, ymax=1.1, ytick = {0.5, 1},
legend image post style={scale=0.4},
legend style={at={(0.995,0.01)},anchor= south east,
nodes={scale=0.4, transform shape}}, grid]
\addplot[matblue, very thick] table[x=rate,y=cdf_rate1] {Data-n3/sr_n3_P15_phi0.90_0.85_0.80_var1.0_2.0_3.0_rate_CDF.txt};
\addplot[matred, very thick] table[x=rate,y=cdf_rate2] {Data-n3/sr_n3_P15_phi0.90_0.85_0.80_var1.0_2.0_3.0_rate_CDF.txt};
\addplot[matorange, very thick] table[x=rate,y=cdf_rate3] {Data-n3/sr_n3_P15_phi0.90_0.85_0.80_var1.0_2.0_3.0_rate_CDF.txt};
\addplot[matblue, very thick, dashed] table[x=rate,y=cdf_rate1] {Data-n3/sr_n3_P15_phi1.00_1.00_1.00_var1.0_2.0_3.0_rate_CDF.txt};
\addplot[matred, very thick, dashed] table[x=rate,y=cdf_rate2] {Data-n3/sr_n3_P15_phi1.00_1.00_1.00_var1.0_2.0_3.0_rate_CDF.txt};
\addplot[matorange, very thick, dashed] table[x=rate,y=cdf_rate3] {Data-n3/sr_n3_P15_phi1.00_1.00_1.00_var1.0_2.0_3.0_rate_CDF.txt};
\legend{$\varphi = 0.90/\sigma^2 = 1.0$, $\varphi = 0.85/\sigma^2 = 2.0$, $\varphi = 0.80/\sigma^2 = 3.0$, $\varphi = 1.00/\sigma^2 = 1.0$, $\varphi = 1.00/\sigma^2 = 2.0$, $\varphi = 1.00/\sigma^2 = 3.0$}
\end{axis}
\end{tikzpicture}
\end{subfigure}
\end{subfigure}
\begin{subfigure}[ht]{.5\linewidth}
\centering
\begin{subfigure}[ht]{\linewidth}
\begin{tikzpicture}[trim axis right,baseline]
\begin{axis}[
width=\linewidth,
height=.4\linewidth,
ylabel={CDF}, 
xmin=0, xmax=9, xmajorticks=false, xtick={0, 3, 6, 9},
ymin=-0.1, ymax=1.1, ytick = {0, 0.5, 1},
legend image post style={scale=0.4},
legend style={at={(0.005,0.985)},anchor= north west,
nodes={scale=0.4, transform shape}}, grid]
\addplot[matblue, very thick] table[x=power,y=cdf_p1] {Data-n3/pf_n3_P15_phi0.90_0.85_0.80_var1.0_2.0_3.0_p_CDF.txt};
\addplot[matred, very thick] table[x=power,y=cdf_p2] {Data-n3/pf_n3_P15_phi0.90_0.85_0.80_var1.0_2.0_3.0_p_CDF.txt};
\addplot[matteal, very thick] table[x=power,y=cdf_p3] {Data-n3/pf_n3_P15_phi0.90_0.85_0.80_var1.0_2.0_3.0_p_CDF.txt};
\addplot[matblue, very thick, dashed] table[x=power,y=cdf_p1] {Data-n3/pf_n3_P15_phi1.00_1.00_1.00_var1.0_2.0_3.0_p_CDF.txt};
\addplot[matred, very thick, dashed] table[x=power,y=cdf_p2] {Data-n3/pf_n3_P15_phi1.00_1.00_1.00_var1.0_2.0_3.0_p_CDF.txt};
\addplot[matorange, very thick, dashed] table[x=power,y=cdf_p3] {Data-n3/pf_n3_P15_phi1.00_1.00_1.00_var1.0_2.0_3.0_p_CDF.txt};
\legend{$\varphi = 0.90/\sigma^2 = 1.0$, $\varphi = 0.85/\sigma^2 = 2.0$, $\varphi = 0.80/\sigma^2 = 3.0$, $\varphi = 1.0/\sigma^2 = 1.0$, $\varphi = 1.0/\sigma^2 = 2.0$, $\varphi = 1.0/\sigma^2 = 3.0$}
\end{axis}
\end{tikzpicture}
\end{subfigure}
\\
\begin{subfigure}[ht]{\linewidth}
\begin{tikzpicture}[trim axis right,baseline]
\begin{axis}[
width=\linewidth,
height=.4\linewidth,
ylabel={CDF}, xlabel={Instantaneous Power}, 
xmin=0, xmax=9, xmajorticks=true, xtick={0, 3, 6, 9},
ymin=-0.1, ymax=1.1, ytick = {0, 0.5, 1},
legend image post style={scale=0.4},
legend style={at={(0.005,0.985)},anchor= north west,
nodes={scale=0.4, transform shape}}, grid]
\addplot[matblue, very thick] table[x=power,y=cdf_p1] {Data-n3/sr_n3_P15_phi0.90_0.85_0.80_var1.0_2.0_3.0_p_CDF.txt};
\addplot[matred, very thick] table[x=power,y=cdf_p2] {Data-n3/sr_n3_P15_phi0.90_0.85_0.80_var1.0_2.0_3.0_p_CDF.txt};
\addplot[matteal, very thick] table[x=power,y=cdf_p3] {Data-n3/sr_n3_P15_phi0.90_0.85_0.80_var1.0_2.0_3.0_p_CDF.txt};
\addplot[matblue, very thick, dashed] table[x=power,y=cdf_p1] {Data-n3/sr_n3_P15_phi1.00_1.00_1.00_var1.0_2.0_3.0_p_CDF.txt};
\addplot[matred, very thick, dashed] table[x=power,y=cdf_p2] {Data-n3/sr_n3_P15_phi1.00_1.00_1.00_var1.0_2.0_3.0_p_CDF.txt};
\addplot[matorange, very thick, dashed] table[x=power,y=cdf_p3] {Data-n3/sr_n3_P15_phi1.00_1.00_1.00_var1.0_2.0_3.0_p_CDF.txt};
\legend{$\varphi = 0.90/\sigma^2 = 1.0$, $\varphi = 0.85/\sigma^2 = 2.0$, $\varphi = 0.80/\sigma^2 = 3.0$, $\varphi = 1.00/\sigma^2 = 1.0$, $\varphi = 1.00/\sigma^2 = 2.0$, $\varphi = 1.00/\sigma^2 = 3.0$}
\end{axis}
\end{tikzpicture}
\end{subfigure}
\end{subfigure}
\caption{Rate (left) and power (right) distributions for a $3$-terminal network with proportional fairness utility (top) and sumrate utility (bottom).}
\label{fig:3_cdfs}
\end{figure*}
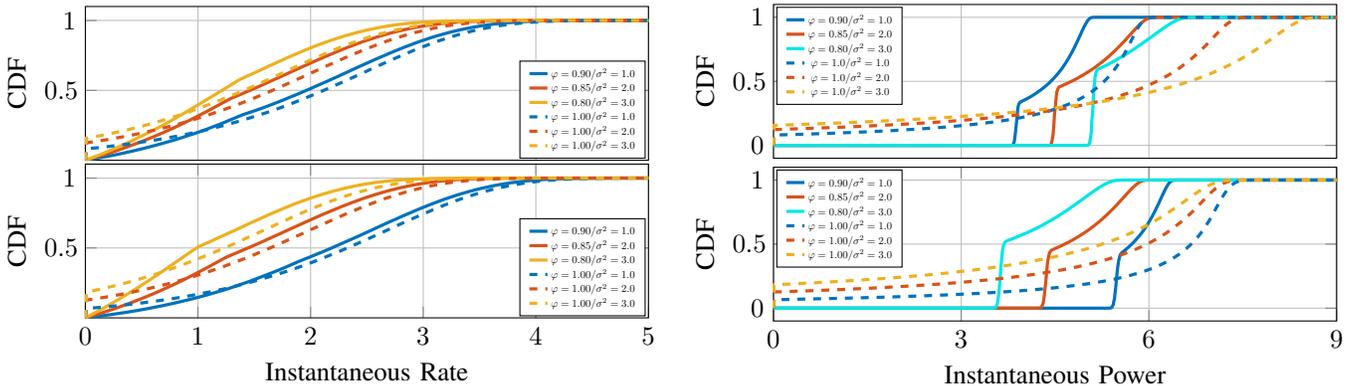

\subsection{Dual Descent}
\label{subsec:dual_descent}

Using the optimal solutions of all primal variables $(\vect{x}, \vect{p}, \vect{z})$, we may express the dual problem as
\begin{multline}
\label{eqn:dual_problem_updated}
D^* = \inf_{\vect{\Lambda}\ \succeq \vect{0}} f_0(\vect{x}^*) + \vect{\lambda}^\top \left( \mathbb{E}\left[ \vect{r}(\vect{p}^*,\cdot) \right] - \vect{x}^* \right) \\
+ \mu  P_0 - \sum_{i=1}^{N_U} \left( z_i^* + \frac{1}{\varphi_i} \mathbb{E}\left[(p_i^* - z_i^*)_+ \right]  \right).
\end{multline}
We utilize the constraint gap to formulate a stochastic subgradient descent update rule for the dual variables $\vect{\Lambda}$ as \cite{ribeiro_optimal_2012}
\begin{equation}
\label{eqn:Lambda_dual_update}
\vect{\Lambda}^{(n)} \triangleq \left( \vect{\Lambda}^{(n-1)} - \varepsilon \Tilde{\vect{g}}_{\vect{\Lambda}}\big( \vect{\Lambda}^{(n-1)} \big) \right)_+, \quad n \geq 0,
\end{equation}
where the stochastic subgradient $\Tilde{\vect{g}}_{\vect{\Lambda}} = ( \Tilde{\vect{g}}_{\vect{\lambda}}, \Tilde{\vect{g}}_{\mu} )^\top$ is
\begin{align*}
\Tilde{\vect{g}}_{\vect{\lambda}}\big( \vect{\Lambda}^{(n-1)} \big) &= \vect{r}\big( \vect{p}^* \big(\vect{\Lambda}^{(n-1)} \big), \vect{h}^{(n)} \big) - \vect{x}^* \big( \vect{\lambda}^{(n-1)} \big), \\
\Tilde{\vect{g}}_{\mu}\big( \vect{\Lambda}^{(n-1)} \big) &= P_0 - \bigg\| \vect{z}^*(\vect{\Lambda}^{(n-1)}) \numberthis \label{eqn:lambda_subgradient}\\
+& \frac{1}{\vect{\varphi}} \odot \left( \vect{p}^*(\vect{\Lambda}^{(n-1)}, \vect{h}^{(n)}) - \vect{z}^*(\vect{\Lambda}^{(n-1)}) \right)_+ \bigg\|_1,
\end{align*}
$\varepsilon$ being the stepsize. Notice that \eqref{eqn:lambda_subgradient} is a stochastic subgradient of the objective of \eqref{eqn:dual_problem_updated} \cite[Theorem~9.108]{shapiro_lectures_2021}. The resulting scheme is summarized in Algorithm~\ref{alg:algorithm_1}. Utilizing the model-free updates in  \eqref{eqn:z_supergradient_ascent} for variables $z_i$ only requires substituting $z_i^*(\vect{\Lambda}^{(n)})$ with iterates $z_i^{(n)}$, as shown in Algorithm~\ref{alg:algorithm_2}.
\vspace{2bp}
\begin{remark}
Note that both Algorithms~\ref{alg:algorithm_1} and~\ref{alg:algorithm_2} are in fact data-driven, since both may be readily implemented based on sampled channel fading data. The difference is that Algorithm~\ref{alg:algorithm_2} can also be implemented \textit{sequentially}.
\end{remark}

\section{Performance Evaluation}
\label{sec:performance_eval}

We now confirm the efficacy of the proposed approach, particularly via Algorithm~\ref{alg:algorithm_1}. We consider $3$-terminal and $8$-terminal point-to-point networks consisting of parallel (independent) links with distinct sets of parameters, e.g., noise variances and CVaR confidence intervals, operating under Rayleigh fading; see Tables \ref{tab:sim_parameters} and \ref{tab:sim_parameters2} for the details. The proposed Algorithm~\ref{alg:algorithm_1} is then applied for the sumrate and proportional fairness utilities; see also Section~\ref{subsec:rate_vector_x}.

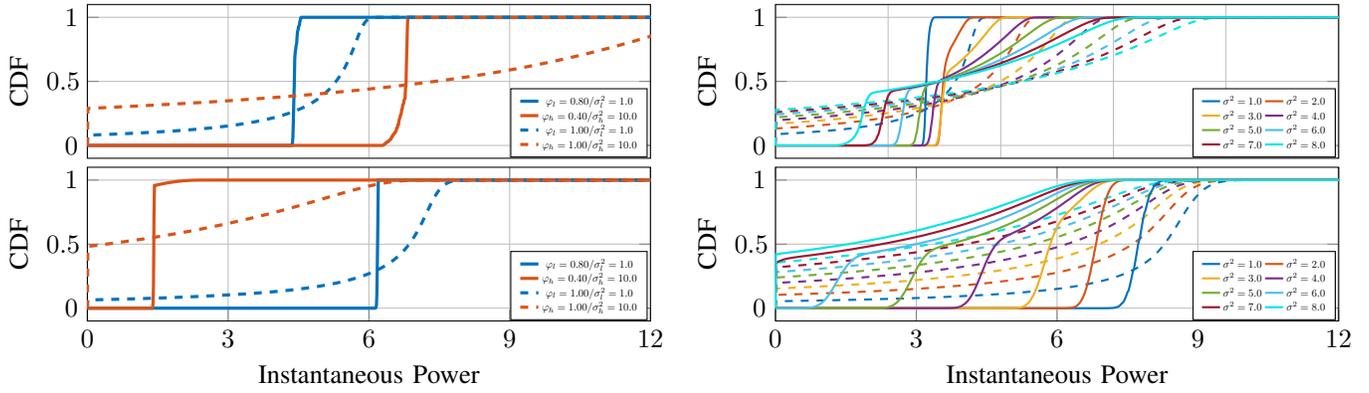
\begin{figure*}[t]
\begin{subfigure}[ht]{.5\linewidth}
\centering
\begin{subfigure}[ht]{\linewidth}
\begin{tikzpicture}[trim axis right,baseline]
\begin{axis}[
width=\linewidth,
height=.4\linewidth,
ylabel={CDF}, 
xmin=0, xmax=12, xmajorticks=false, xtick={0, 3, 6, 9, 12, 15},
ymin=-0.1, ymax=1.1, ytick = {0, 0.5, 1},
legend image post style={scale=0.4},
legend style={at={(0.995,0.005)},anchor= south east,
nodes={scale=0.4, transform shape}}, grid]
\addplot[matblue, very thick] table[x=power,y=cdf_p1] {Data-n8/pf_n8_P40_phi0.80_0.80_0.80_0.80_0.80_0.80_0.40_0.40_var1.0_1.0_1.0_1.0_1.0_1.0_10.0_10.0_p_CDF.txt};
\addplot[matred, very thick] table[x=power,y=cdf_p7] {Data-n8/pf_n8_P40_phi0.80_0.80_0.80_0.80_0.80_0.80_0.40_0.40_var1.0_1.0_1.0_1.0_1.0_1.0_10.0_10.0_p_CDF.txt};
\addplot[matblue, very thick, dashed] table[x=power,y=cdf_p1] {Data-n8/pf_n8_P40_phi1.00_1.00_1.00_1.00_1.00_1.00_1.00_1.00_var1.0_1.0_1.0_1.0_1.0_1.0_10.0_10.0_p_CDF.txt};
\addplot[matred, very thick, dashed] table[x=power,y=cdf_p7] {Data-n8/pf_n8_P40_phi1.00_1.00_1.00_1.00_1.00_1.00_1.00_1.00_var1.0_1.0_1.0_1.0_1.0_1.0_10.0_10.0_p_CDF.txt};

\legend{$\varphi_l = 0.80/\sigma_l^2 = 1.0$, $\varphi_h = 0.40/\sigma_h^2 = 10.0$, $\varphi_l = 1.00/\sigma_l^2 = 1.0$, $\varphi_h = 1.00/\sigma_h^2 = 10.0$}
\end{axis}
\end{tikzpicture}
\end{subfigure}
\begin{subfigure}[ht]{\linewidth}
\begin{tikzpicture}[trim axis right,baseline]
\begin{axis}[
width=\linewidth,
height=.4\linewidth,
ylabel={CDF}, xlabel={Instantaneous Power},
xmin=0, xmax=12, xmajorticks=true, xtick={0, 3, 6, 9, 12, 15},
ymin=-0.1, ymax=1.1, ytick = {0, 0.5, 1},
legend image post style={scale=0.4},
legend style={at={(0.995,0.005)},anchor= south east,
nodes={scale=0.4, transform shape}}, grid]
\addplot[matblue, very thick] table[x=power,y=cdf_p1] {Data-n8/sr_n8_P40_phi0.80_0.80_0.80_0.80_0.80_0.80_0.40_0.40_var1.0_1.0_1.0_1.0_1.0_1.0_10.0_10.0_p_CDF.txt};
\addplot[matred, very thick] table[x=power,y=cdf_p7] {Data-n8/sr_n8_P40_phi0.80_0.80_0.80_0.80_0.80_0.80_0.40_0.40_var1.0_1.0_1.0_1.0_1.0_1.0_10.0_10.0_p_CDF.txt};
\addplot[matblue, very thick, dashed] table[x=power,y=cdf_p1] {Data-n8/sr_n8_P40_phi1.00_1.00_1.00_1.00_1.00_1.00_1.00_1.00_var1.0_1.0_1.0_1.0_1.0_1.0_10.0_10.0_p_CDF.txt};
\addplot[matred, very thick, dashed] table[x=power,y=cdf_p7] {Data-n8/sr_n8_P40_phi1.00_1.00_1.00_1.00_1.00_1.00_1.00_1.00_var1.0_1.0_1.0_1.0_1.0_1.0_10.0_10.0_p_CDF.txt};

\legend{$\varphi_l = 0.80/\sigma_l^2 = 1.0$, $\varphi_h = 0.40/\sigma_h^2 = 10.0$, $\varphi_l = 1.00/\sigma_l^2 = 1.0$, $\varphi_h = 1.00/\sigma_h^2 = 10.0$}
\end{axis}
\end{tikzpicture}
\end{subfigure}
\end{subfigure}
\begin{subfigure}[ht]{.5\linewidth}
\centering
\begin{subfigure}[ht]{\linewidth}
\begin{tikzpicture}[trim axis right,baseline]
\begin{axis}[
width=\linewidth,
height=.4\linewidth,
ylabel={CDF}, 
xmin=0, xmax=15, xmajorticks=false, xtick={0, 3, 6, 9, 12, 15},
ymin=-0.1, ymax=1.1, ytick = {0, 0.5, 1},
legend image post style={scale=0.4},
legend style={at={(0.995,0.005)},anchor= south east,
nodes={scale=0.4, transform shape}}, legend columns = 2, grid]
\addplot[matblue, thick, dashed] table[x=power,y=cdf_p1] {Data-n8/pf_n8_P40_phi1.00_1.00_1.00_1.00_1.00_1.00_1.00_1.00_var1.0_2.0_3.0_4.0_5.0_6.0_7.0_8.0_p_CDF.txt};
\addplot[matred, thick, dashed] table[x=power,y=cdf_p2] {Data-n8/pf_n8_P40_phi1.00_1.00_1.00_1.00_1.00_1.00_1.00_1.00_var1.0_2.0_3.0_4.0_5.0_6.0_7.0_8.0_p_CDF.txt};
\addplot[matorange, thick, dashed] table[x=power,y=cdf_p3] {Data-n8/pf_n8_P40_phi1.00_1.00_1.00_1.00_1.00_1.00_1.00_1.00_var1.0_2.0_3.0_4.0_5.0_6.0_7.0_8.0_p_CDF.txt};
\addplot[matviolet, thick, dashed] table[x=power,y=cdf_p4] {Data-n8/pf_n8_P40_phi1.00_1.00_1.00_1.00_1.00_1.00_1.00_1.00_var1.0_2.0_3.0_4.0_5.0_6.0_7.0_8.0_p_CDF.txt};
\addplot[matgreen, thick, dashed] table[x=power,y=cdf_p5] {Data-n8/pf_n8_P40_phi1.00_1.00_1.00_1.00_1.00_1.00_1.00_1.00_var1.0_2.0_3.0_4.0_5.0_6.0_7.0_8.0_p_CDF.txt};
\addplot[matskyblue, thick, dashed] table[x=power,y=cdf_p6] {Data-n8/pf_n8_P40_phi1.00_1.00_1.00_1.00_1.00_1.00_1.00_1.00_var1.0_2.0_3.0_4.0_5.0_6.0_7.0_8.0_p_CDF.txt};
\addplot[matburgundy, thick, dashed] table[x=power,y=cdf_p7] {Data-n8/pf_n8_P40_phi1.00_1.00_1.00_1.00_1.00_1.00_1.00_1.00_var1.0_2.0_3.0_4.0_5.0_6.0_7.0_8.0_p_CDF.txt};
\addplot[matteal, thick, dashed] table[x=power,y=cdf_p8] {Data-n8/pf_n8_P40_phi1.00_1.00_1.00_1.00_1.00_1.00_1.00_1.00_var1.0_2.0_3.0_4.0_5.0_6.0_7.0_8.0_p_CDF.txt};
\addplot[matblue, thick] table[x=power,y=cdf_p1] {Data-n8/pf_n8_P40_phi0.80_0.80_0.80_0.80_0.80_0.80_0.80_0.80_var1.0_2.0_3.0_4.0_5.0_6.0_7.0_8.0_p_CDF.txt};
\addplot[matred, thick] table[x=power,y=cdf_p2] {Data-n8/pf_n8_P40_phi0.80_0.80_0.80_0.80_0.80_0.80_0.80_0.80_var1.0_2.0_3.0_4.0_5.0_6.0_7.0_8.0_p_CDF.txt};
\addplot[matorange, thick] table[x=power,y=cdf_p3] {Data-n8/pf_n8_P40_phi0.80_0.80_0.80_0.80_0.80_0.80_0.80_0.80_var1.0_2.0_3.0_4.0_5.0_6.0_7.0_8.0_p_CDF.txt};
\addplot[matviolet, thick] table[x=power,y=cdf_p4] {Data-n8/pf_n8_P40_phi0.80_0.80_0.80_0.80_0.80_0.80_0.80_0.80_var1.0_2.0_3.0_4.0_5.0_6.0_7.0_8.0_p_CDF.txt};
\addplot[matgreen, thick] table[x=power,y=cdf_p5] {Data-n8/pf_n8_P40_phi0.80_0.80_0.80_0.80_0.80_0.80_0.80_0.80_var1.0_2.0_3.0_4.0_5.0_6.0_7.0_8.0_p_CDF.txt};
\addplot[matskyblue, thick] table[x=power,y=cdf_p6] {Data-n8/pf_n8_P40_phi0.80_0.80_0.80_0.80_0.80_0.80_0.80_0.80_var1.0_2.0_3.0_4.0_5.0_6.0_7.0_8.0_p_CDF.txt};
\addplot[matburgundy, thick] table[x=power,y=cdf_p7] {Data-n8/pf_n8_P40_phi0.80_0.80_0.80_0.80_0.80_0.80_0.80_0.80_var1.0_2.0_3.0_4.0_5.0_6.0_7.0_8.0_p_CDF.txt};
\addplot[matteal, thick] table[x=power,y=cdf_p8] {Data-n8/pf_n8_P40_phi0.80_0.80_0.80_0.80_0.80_0.80_0.80_0.80_var1.0_2.0_3.0_4.0_5.0_6.0_7.0_8.0_p_CDF.txt};
\legend{ , , , , , , , , $\sigma^2 = 1.0$, $\sigma^2 = 2.0$, $\sigma^2 = 3.0$, $\sigma^2 = 4.0$, $\sigma^2 = 5.0$, $\sigma^2 = 6.0$, $\sigma^2 = 7.0$, $\sigma^2 = 8.0$}
\end{axis}
\end{tikzpicture}
\end{subfigure}
\begin{subfigure}[ht]{\linewidth}
\begin{tikzpicture}[trim axis right,baseline]
\begin{axis}[
width=\linewidth,
height=.4\linewidth,
ylabel={CDF}, xlabel={Instantaneous Power},
xmin=0, xmax=12, xmajorticks=true, xtick={0, 3, 6, 9, 12},
ymin=-0.1, ymax=1.1, ytick = {0, 0.5, 1},
legend image post style={scale=0.4},
legend style={at={(0.995,0.005)},anchor= south east,
nodes={scale=0.4, transform shape}}, legend columns = 2, grid]
\addplot[matblue, thick, dashed] table[x=power,y=cdf_p1] {Data-n8/sr_n8_P40_phi1.00_1.00_1.00_1.00_1.00_1.00_1.00_1.00_var1.0_2.0_3.0_4.0_5.0_6.0_7.0_8.0_p_CDF.txt};
\addplot[matred, thick, dashed] table[x=power,y=cdf_p2] {Data-n8/sr_n8_P40_phi1.00_1.00_1.00_1.00_1.00_1.00_1.00_1.00_var1.0_2.0_3.0_4.0_5.0_6.0_7.0_8.0_p_CDF.txt};
\addplot[matorange, thick, dashed] table[x=power,y=cdf_p3] {Data-n8/sr_n8_P40_phi1.00_1.00_1.00_1.00_1.00_1.00_1.00_1.00_var1.0_2.0_3.0_4.0_5.0_6.0_7.0_8.0_p_CDF.txt};
\addplot[matviolet, thick, dashed] table[x=power,y=cdf_p4] {Data-n8/sr_n8_P40_phi1.00_1.00_1.00_1.00_1.00_1.00_1.00_1.00_var1.0_2.0_3.0_4.0_5.0_6.0_7.0_8.0_p_CDF.txt};
\addplot[matgreen, thick, dashed] table[x=power,y=cdf_p5] {Data-n8/sr_n8_P40_phi1.00_1.00_1.00_1.00_1.00_1.00_1.00_1.00_var1.0_2.0_3.0_4.0_5.0_6.0_7.0_8.0_p_CDF.txt};
\addplot[matskyblue, thick, dashed] table[x=power,y=cdf_p6] {Data-n8/sr_n8_P40_phi1.00_1.00_1.00_1.00_1.00_1.00_1.00_1.00_var1.0_2.0_3.0_4.0_5.0_6.0_7.0_8.0_p_CDF.txt};
\addplot[matburgundy, thick, dashed] table[x=power,y=cdf_p7] {Data-n8/sr_n8_P40_phi1.00_1.00_1.00_1.00_1.00_1.00_1.00_1.00_var1.0_2.0_3.0_4.0_5.0_6.0_7.0_8.0_p_CDF.txt};
\addplot[matteal, thick, dashed] table[x=power,y=cdf_p8] {Data-n8/sr_n8_P40_phi1.00_1.00_1.00_1.00_1.00_1.00_1.00_1.00_var1.0_2.0_3.0_4.0_5.0_6.0_7.0_8.0_p_CDF.txt};
\addplot[matblue, thick] table[x=power,y=cdf_p1] {Data-n8/sr_n8_P40_phi0.80_0.80_0.80_0.80_0.80_0.80_0.80_0.80_var1.0_2.0_3.0_4.0_5.0_6.0_7.0_8.0_p_CDF.txt};
\addplot[matred, thick] table[x=power,y=cdf_p2] {Data-n8/sr_n8_P40_phi0.80_0.80_0.80_0.80_0.80_0.80_0.80_0.80_var1.0_2.0_3.0_4.0_5.0_6.0_7.0_8.0_p_CDF.txt};
\addplot[matorange, thick] table[x=power,y=cdf_p3] {Data-n8/sr_n8_P40_phi0.80_0.80_0.80_0.80_0.80_0.80_0.80_0.80_var1.0_2.0_3.0_4.0_5.0_6.0_7.0_8.0_p_CDF.txt};
\addplot[matviolet, thick] table[x=power,y=cdf_p4] {Data-n8/sr_n8_P40_phi0.80_0.80_0.80_0.80_0.80_0.80_0.80_0.80_var1.0_2.0_3.0_4.0_5.0_6.0_7.0_8.0_p_CDF.txt};
\addplot[matgreen, thick] table[x=power,y=cdf_p5] {Data-n8/sr_n8_P40_phi0.80_0.80_0.80_0.80_0.80_0.80_0.80_0.80_var1.0_2.0_3.0_4.0_5.0_6.0_7.0_8.0_p_CDF.txt};
\addplot[matskyblue, thick] table[x=power,y=cdf_p6] {Data-n8/sr_n8_P40_phi0.80_0.80_0.80_0.80_0.80_0.80_0.80_0.80_var1.0_2.0_3.0_4.0_5.0_6.0_7.0_8.0_p_CDF.txt};
\addplot[matburgundy, thick] table[x=power,y=cdf_p7] {Data-n8/sr_n8_P40_phi0.80_0.80_0.80_0.80_0.80_0.80_0.80_0.80_var1.0_2.0_3.0_4.0_5.0_6.0_7.0_8.0_p_CDF.txt};
\addplot[matteal, thick] table[x=power,y=cdf_p8] {Data-n8/sr_n8_P40_phi0.80_0.80_0.80_0.80_0.80_0.80_0.80_0.80_var1.0_2.0_3.0_4.0_5.0_6.0_7.0_8.0_p_CDF.txt};
\legend{ , , , , , , , , $\sigma^2 = 1.0$, $\sigma^2 = 2.0$, $\sigma^2 = 3.0$, $\sigma^2 = 4.0$, $\sigma^2 = 5.0$, $\sigma^2 = 6.0$, $\sigma^2 = 7.0$, $\sigma^2 = 8.0$}
\end{axis}
\end{tikzpicture}
\end{subfigure}
\end{subfigure}
\caption{Optimal power distributions (CDFs) for realistic (left) and toy example (right) $8$-terminal networks with proportional fairness utility (top) and sumrate utility (bottom).}
\label{fig:p8_cdf}
\end{figure*}
\begin{figure*}[t]
\begin{subfigure}[ht]{.5\linewidth}
\centering
\begin{subfigure}[ht]{\linewidth}
\begin{tikzpicture}[trim axis right,baseline]
\begin{axis}[
width=\linewidth,
height=.45\linewidth,
ytick={0.5, 1, 1.5, 2, 2.5},
ylabel={Rate},
xmin=1, xmax=500, ymin=0, ymax=3.1, xmajorticks=false,
legend style={at={(1,1)},anchor=north east,
nodes={scale=0.4, transform shape}, legend columns = 4},
grid]
\addplot[matblue, thick, mark=no] table[x=time_r,y=cum_r1] {Data-n8/pf_n8_P40_phi0.40_0.40_0.40_0.40_0.40_0.40_0.80_0.80_var1.0_1.0_1.0_1.0_1.0_1.0_10.0_10.0_rate_instances.txt};
\addplot[matorange, thick, mark=no] table[x=time_r,y=cum_r2] {Data-n8/pf_n8_P40_phi0.40_0.40_0.40_0.40_0.40_0.40_0.80_0.80_var1.0_1.0_1.0_1.0_1.0_1.0_10.0_10.0_rate_instances.txt};
\addplot[matgreen, thick, mark=no] table[x=time_r,y=cum_r3] {Data-n8/pf_n8_P40_phi0.40_0.40_0.40_0.40_0.40_0.40_0.80_0.80_var1.0_1.0_1.0_1.0_1.0_1.0_10.0_10.0_rate_instances.txt};
\addplot[matskyblue, thick, mark=no] table[x=time_r,y=cum_r4] {Data-n8/pf_n8_P40_phi0.40_0.40_0.40_0.40_0.40_0.40_0.80_0.80_var1.0_1.0_1.0_1.0_1.0_1.0_10.0_10.0_rate_instances.txt};
\addplot[matviolet, thick, mark=no] table[x=time_r,y=cum_r5] {Data-n8/pf_n8_P40_phi0.40_0.40_0.40_0.40_0.40_0.40_0.80_0.80_var1.0_1.0_1.0_1.0_1.0_1.0_10.0_10.0_rate_instances.txt};
\addplot[matteal, thick, mark=no] table[x=time_r,y=cum_r6] {Data-n8/pf_n8_P40_phi0.40_0.40_0.40_0.40_0.40_0.40_0.80_0.80_var1.0_1.0_1.0_1.0_1.0_1.0_10.0_10.0_rate_instances.txt};
\addplot[matred, thick, mark=no] table[x=time_r,y=cum_r7] {Data-n8/pf_n8_P40_phi0.40_0.40_0.40_0.40_0.40_0.40_0.80_0.80_var1.0_1.0_1.0_1.0_1.0_1.0_10.0_10.0_rate_instances.txt};
\addplot[matburgundy, thick, mark=no] table[x=time_r,y=cum_r8] {Data-n8/pf_n8_P40_phi0.40_0.40_0.40_0.40_0.40_0.40_0.80_0.80_var1.0_1.0_1.0_1.0_1.0_1.0_10.0_10.0_rate_instances.txt};
\addplot[black, very thick, mark=no] table[x=time_r,y=sum_cum_r] {Data-n8/pf_n8_P40_phi0.40_0.40_0.40_0.40_0.40_0.40_0.80_0.80_var1.0_1.0_1.0_1.0_1.0_1.0_10.0_10.0_rate_instances.txt};
\addplot[black, very thick, mark=no, dashed] table[x=time_r,y=sum_cum_r] {Data-n8/pf_n8_P40_phi1.00_1.00_1.00_1.00_1.00_1.00_1.00_1.00_var1.0_1.0_1.0_1.0_1.0_1.0_10.0_10.0_rate_instances.txt};
\legend{$\varphi_l = 0.80/\sigma_l^2 = 1.0$, , , , , , $\varphi_h = 0.40/\sigma_h^2 = 10.0$, , Avg, RN-Avg}
\end{axis}
\end{tikzpicture}
\end{subfigure}
\begin{subfigure}[ht]{\linewidth}
\begin{tikzpicture}[trim axis right,baseline]
\begin{axis}[
width=\linewidth,
height=.45\linewidth,
ytick={0.5, 1, 1.5, 2, 2.5},
ylabel={Rate},
xlabel={Time (Fading Realizations)},
xmin=1, xmax=500, ymin=0, ymax=3.1, xmajorticks=true,
legend style={at={(1,1)},anchor=north east,
nodes={scale=0.4, transform shape}, legend columns = 4},
grid]
\addplot[matblue, thick, mark=no] table[x=time_r,y=cum_r1] {Data-n8/sr_n8_P40_phi0.40_0.40_0.40_0.40_0.40_0.40_0.80_0.80_var1.0_1.0_1.0_1.0_1.0_1.0_10.0_10.0_rate_instances.txt};
\addplot[matorange, thick, mark=no] table[x=time_r,y=cum_r2] {Data-n8/sr_n8_P40_phi0.40_0.40_0.40_0.40_0.40_0.40_0.80_0.80_var1.0_1.0_1.0_1.0_1.0_1.0_10.0_10.0_rate_instances.txt};
\addplot[matgreen, thick, mark=no] table[x=time_r,y=cum_r3] {Data-n8/sr_n8_P40_phi0.40_0.40_0.40_0.40_0.40_0.40_0.80_0.80_var1.0_1.0_1.0_1.0_1.0_1.0_10.0_10.0_rate_instances.txt};
\addplot[matskyblue, thick, mark=no] table[x=time_r,y=cum_r4] {Data-n8/sr_n8_P40_phi0.40_0.40_0.40_0.40_0.40_0.40_0.80_0.80_var1.0_1.0_1.0_1.0_1.0_1.0_10.0_10.0_rate_instances.txt};
\addplot[matviolet, thick, mark=no] table[x=time_r,y=cum_r5] {Data-n8/sr_n8_P40_phi0.40_0.40_0.40_0.40_0.40_0.40_0.80_0.80_var1.0_1.0_1.0_1.0_1.0_1.0_10.0_10.0_rate_instances.txt};
\addplot[matteal, thick, mark=no] table[x=time_r,y=cum_r6] {Data-n8/sr_n8_P40_phi0.40_0.40_0.40_0.40_0.40_0.40_0.80_0.80_var1.0_1.0_1.0_1.0_1.0_1.0_10.0_10.0_rate_instances.txt};
\addplot[matred, thick, mark=no] table[x=time_r,y=cum_r7] {Data-n8/sr_n8_P40_phi0.40_0.40_0.40_0.40_0.40_0.40_0.80_0.80_var1.0_1.0_1.0_1.0_1.0_1.0_10.0_10.0_rate_instances.txt};
\addplot[matburgundy, thick, mark=no] table[x=time_r,y=cum_r8] {Data-n8/sr_n8_P40_phi0.40_0.40_0.40_0.40_0.40_0.40_0.80_0.80_var1.0_1.0_1.0_1.0_1.0_1.0_10.0_10.0_rate_instances.txt};
\addplot[black, very thick, mark=no] table[x=time_r,y=sum_cum_r] {Data-n8/sr_n8_P40_phi0.40_0.40_0.40_0.40_0.40_0.40_0.80_0.80_var1.0_1.0_1.0_1.0_1.0_1.0_10.0_10.0_rate_instances.txt};
\addplot[black, very thick, mark=no, dashed] table[x=time_r,y=sum_cum_r] {Data-n8/sr_n8_P40_phi1.00_1.00_1.00_1.00_1.00_1.00_1.00_1.00_var1.0_1.0_1.0_1.0_1.0_1.0_10.0_10.0_rate_instances.txt};
\legend{$\varphi_l = 0.40/\sigma_l^2 = 1.0$, , , , , , $\varphi_h = 0.80/\sigma_h^2 = 10.0$, , Avg, RN-Avg}
\end{axis}
\end{tikzpicture}
\end{subfigure}
\end{subfigure}
\begin{subfigure}[ht]{.5\linewidth}
\centering
\begin{subfigure}[ht]{\linewidth}
\begin{tikzpicture}[trim axis right,baseline]
\begin{axis}[
width=\linewidth,
height=.45\linewidth,
ytick={0.5, 1, 1.5, 2, 2.5},
ylabel={Rate},
xmin=1, xmax=500, ymin=0, ymax=3, xmajorticks=false,
legend style={at={(0,1)},anchor=north west,
nodes={scale=0.4, transform shape}, legend columns = 5},
grid]
\addplot[matblue, thick, mark=no] table[x=time_r,y=cum_r1] {Data-n8/pf_n8_P40_phi0.80_0.80_0.80_0.80_0.80_0.80_0.80_0.80_var1.0_2.0_3.0_4.0_5.0_6.0_7.0_8.0_rate_instances.txt};
\addplot[matred, thick, mark=no] table[x=time_r,y=cum_r2] {Data-n8/pf_n8_P40_phi0.80_0.80_0.80_0.80_0.80_0.80_0.80_0.80_var1.0_2.0_3.0_4.0_5.0_6.0_7.0_8.0_rate_instances.txt};
\addplot[matorange, thick, mark=no] table[x=time_r,y=cum_r3] {Data-n8/pf_n8_P40_phi0.80_0.80_0.80_0.80_0.80_0.80_0.80_0.80_var1.0_2.0_3.0_4.0_5.0_6.0_7.0_8.0_rate_instances.txt};
\addplot[matviolet, thick, mark=no] table[x=time_r,y=cum_r4] {Data-n8/pf_n8_P40_phi0.80_0.80_0.80_0.80_0.80_0.80_0.80_0.80_var1.0_2.0_3.0_4.0_5.0_6.0_7.0_8.0_rate_instances.txt};
\addplot[matgreen, thick, mark=no] table[x=time_r,y=cum_r5] {Data-n8/pf_n8_P40_phi0.80_0.80_0.80_0.80_0.80_0.80_0.80_0.80_var1.0_2.0_3.0_4.0_5.0_6.0_7.0_8.0_rate_instances.txt};
\addplot[matskyblue, thick, mark=no] table[x=time_r,y=cum_r6] {Data-n8/pf_n8_P40_phi0.80_0.80_0.80_0.80_0.80_0.80_0.80_0.80_var1.0_2.0_3.0_4.0_5.0_6.0_7.0_8.0_rate_instances.txt};
\addplot[matburgundy, thick, mark=no] table[x=time_r,y=cum_r7] {Data-n8/pf_n8_P40_phi0.80_0.80_0.80_0.80_0.80_0.80_0.80_0.80_var1.0_2.0_3.0_4.0_5.0_6.0_7.0_8.0_rate_instances.txt};
\addplot[matteal, thick, mark=no] table[x=time_r,y=cum_r8] {Data-n8/pf_n8_P40_phi0.80_0.80_0.80_0.80_0.80_0.80_0.80_0.80_var1.0_2.0_3.0_4.0_5.0_6.0_7.0_8.0_rate_instances.txt};
\addplot[black, very thick, mark=no] table[x=time_r,y=sum_cum_r] {Data-n8/pf_n8_P40_phi0.80_0.80_0.80_0.80_0.80_0.80_0.80_0.80_var1.0_2.0_3.0_4.0_5.0_6.0_7.0_8.0_rate_instances.txt};
\legend{$\sigma^2 = 1.0$, $\sigma^2 = 2.0$, $\sigma^2 = 3.0$, $\sigma^2 = 4.0$, $\sigma^2 = 5.0$, $\sigma^2 = 6.0$, $\sigma^2 = 7.0$, $\sigma^2 = 8.0$, Avg}
\end{axis}
\end{tikzpicture}
\end{subfigure}
\begin{subfigure}[ht]{\linewidth}
\begin{tikzpicture}[trim axis right,baseline]
\begin{axis}[
width=\linewidth,
height=.45\linewidth,
ytick={0.5, 1, 1.5, 2, 2.5},
ylabel={Rate},
xlabel={Time (Fading Realizations)},
xmin=1, xmax=500, ymin=0, ymax=3, xmajorticks=true,
legend style={at={(0,1)},anchor=north west,
nodes={scale=0.4, transform shape}, legend columns = 5},
grid]
\addplot[matblue, thick, mark=no] table[x=time_r,y=cum_r1] {Data-n8/sr_n8_P40_phi0.80_0.80_0.80_0.80_0.80_0.80_0.80_0.80_var1.0_2.0_3.0_4.0_5.0_6.0_7.0_8.0_rate_instances.txt};
\addplot[matred, thick, mark=no] table[x=time_r,y=cum_r2] {Data-n8/sr_n8_P40_phi0.80_0.80_0.80_0.80_0.80_0.80_0.80_0.80_var1.0_2.0_3.0_4.0_5.0_6.0_7.0_8.0_rate_instances.txt};
\addplot[matorange, thick, mark=no] table[x=time_r,y=cum_r3] {Data-n8/sr_n8_P40_phi0.80_0.80_0.80_0.80_0.80_0.80_0.80_0.80_var1.0_2.0_3.0_4.0_5.0_6.0_7.0_8.0_rate_instances.txt};
\addplot[matviolet, thick, mark=no] table[x=time_r,y=cum_r4] {Data-n8/sr_n8_P40_phi0.80_0.80_0.80_0.80_0.80_0.80_0.80_0.80_var1.0_2.0_3.0_4.0_5.0_6.0_7.0_8.0_rate_instances.txt};
\addplot[matgreen, thick, mark=no] table[x=time_r,y=cum_r5] {Data-n8/sr_n8_P40_phi0.80_0.80_0.80_0.80_0.80_0.80_0.80_0.80_var1.0_2.0_3.0_4.0_5.0_6.0_7.0_8.0_rate_instances.txt};
\addplot[matskyblue, thick, mark=no] table[x=time_r,y=cum_r6] {Data-n8/sr_n8_P40_phi0.80_0.80_0.80_0.80_0.80_0.80_0.80_0.80_var1.0_2.0_3.0_4.0_5.0_6.0_7.0_8.0_rate_instances.txt};
\addplot[matburgundy, thick, mark=no] table[x=time_r,y=cum_r7] {Data-n8/sr_n8_P40_phi0.80_0.80_0.80_0.80_0.80_0.80_0.80_0.80_var1.0_2.0_3.0_4.0_5.0_6.0_7.0_8.0_rate_instances.txt};
\addplot[matteal, thick, mark=no] table[x=time_r,y=cum_r8] {Data-n8/sr_n8_P40_phi0.80_0.80_0.80_0.80_0.80_0.80_0.80_0.80_var1.0_2.0_3.0_4.0_5.0_6.0_7.0_8.0_rate_instances.txt};
\addplot[black, very thick, mark=no] table[x=time_r,y=sum_cum_r] {Data-n8/sr_n8_P40_phi0.80_0.80_0.80_0.80_0.80_0.80_0.80_0.80_var1.0_2.0_3.0_4.0_5.0_6.0_7.0_8.0_rate_instances.txt};
\legend{$\sigma^2 = 1.0$, $\sigma^2 = 2.0$, $\sigma^2 = 3.0$, $\sigma^2 = 4.0$, $\sigma^2 = 5.0$, $\sigma^2 = 6.0$, $\sigma^2 = 7.0$, $\sigma^2 = 8.0$, Avg}
\end{axis}
\end{tikzpicture}
\end{subfigure}
\end{subfigure}
\caption{Moving average of transmission rates of realistic example (left) and toy example (right) $8$-terminal networks in learning period for proportional fairness utility (top) and sumrate utility (bottom). ``RN" stands for ``risk-neutral".}
\label{fig:rate_cummean}
\end{figure*}
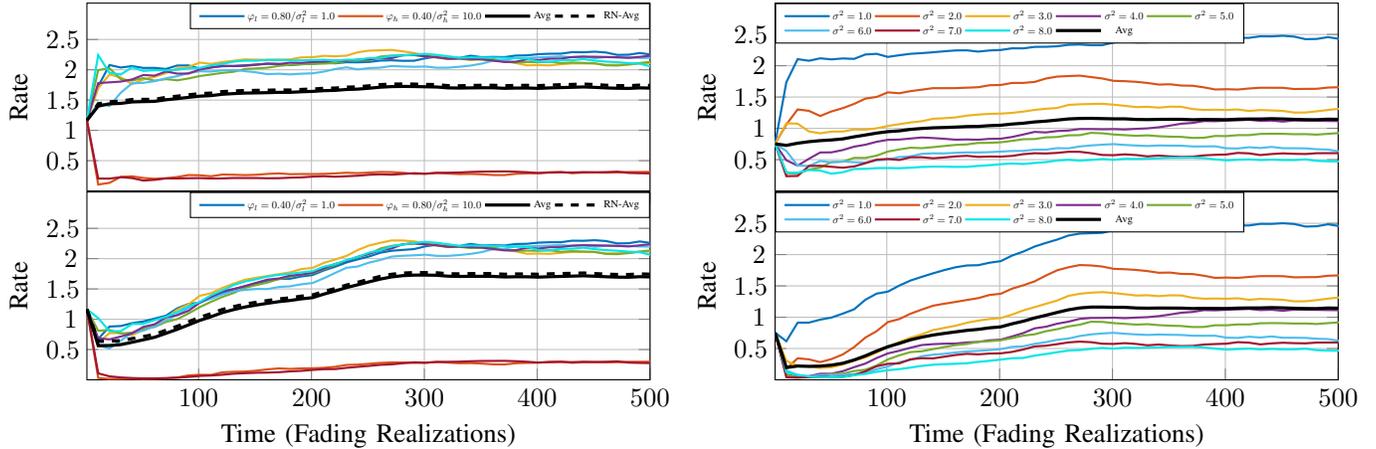

\begin{remark}
Since both Algorithm~\ref{alg:algorithm_1} and \ref{alg:algorithm_2} are data-driven, we find it sufficient to empirically evaluate Algorithm~\ref{alg:algorithm_1}, since the only main difference between the two algorithms is the computational complexity in the optimization of the CVaR quantile levels $\vect{z}$ (rendering Algorithm \ref{alg:algorithm_2} somewhat slower).
\end{remark}

First, we observe that CVaR-optimal policies exhibit a sharp statistical threshold corresponding to the CVaR quantile level variable $\vect{z}$, as shown in Figs.~\ref{fig:3_cdfs}(right) and \ref{fig:p8_cdf}, due to the distributionally robust / risk-averse characteristics of CVaR-optimal resource allocation (problem \eqref{eqn:problem_formulation_RA}). For smaller values of the confidence level $\varphi$, 
implying stricter power regulation,
the variable $\vect{z}$ prevails producing more robust and nearly constant power policies, ensuring increased statistical stability in both individual and total power allocation. As shown in Fig.~\ref{fig:p_vals}, power policy variability is indeed drastically reduced in the case of distributionally robust allocation, compared with its risk-neutral counterpart (i.e., $\vect{\varphi}=\mathbf{1}$). For larger values of $\varphi$, policy selection is relaxed and total variation is increased. 

\begin{table}[t]
\vspace{8pt}
\centering
\caption{Simulation parameters for $3$-terminal network}
\begin{tabular}{c||c}
\toprule
$\vect{w}$ & $\begin{pmatrix} \sfrac{1}{3} & \sfrac{1}{3} & \sfrac{1}{3} \end{pmatrix}^\top$\\
$\vect{\sigma}^2$ & $\begin{pmatrix} 1.0 & 2.0 & 3.0 \end{pmatrix}^\top$\\
$\vect{\varphi}$ & $\begin{pmatrix} 0.90 & 0.85 & 0.80 \end{pmatrix}^\top$\\
$(P_0, \varepsilon) $ & $\left( 15, 3 \times 10^{-5} \right)$\\
\bottomrule
\end{tabular}
\label{tab:sim_parameters}
\end{table}

\begin{table}[t]
\centering
\caption{Simulation parameters of $8$-terminal networks}
\begin{tabular}{c||c|c||c}
\multicolumn{2}{c|}{Toy example} & \multicolumn{2}{c}{Realistic example}\\
\toprule
$\vect{w}$ & $(\sfrac{1}{8}) \cdot \vect{1}^\top$ & $\vect{w}$ & $(\sfrac{1}{8}) \cdot \vect{1}^\top$\\
$\vect{\sigma}^2$ & $\begin{pmatrix} 1.0 & \dots & 8.0 \end{pmatrix}^\top$ & $(\sigma_l^2, \sigma_h^2)$ & $(1.0, 10.0)$\\
$\vect{\varphi}$ & $(0.8) \cdot \vect{1}^\top$ & $(\varphi_l, \varphi_h)$ & $(0.40, 0.80)$\\
$(P_0, \varepsilon) $ & $\left( 40, 3\times 10^{-5} \right)$ & $(P_0, \varepsilon) $ & $\left( 40, 3\times 10^{-5} \right)$\\
Window & $200$ & Window & $200$\\
\bottomrule
\end{tabular}
\label{tab:sim_parameters2}
\vspace{4pt}
\end{table}

Quite interestingly, a small confidence level $\varphi$ carries little impact on the achieved ergodic rates, those primarily depending on the noise variances. Smaller values of $\varphi$ induce slightly decreased average transmission rate as compared with risk-neutral counterpart (as expected), as the optimal power policies vary in a narrower set of values ---nearly constant in practice. 
Still, the rate loss is negligibly small; see Figs.~\ref{fig:rate_cummean} and \ref{fig:3_cdfs}(left).


The effect of channel noise variance on system performance is expectedly significant. The transmission rate is inversely proportional to noise variance; see \eqref{eqn:rate_function}. However, the effect depends on the selection of the utility function $f_0$. For the proportional fairness utility (see Section~\ref{subsec:rate_vector_x}), our algorithm maximizes the transmission rates while aiming to be fair among terminals, thus the optimal policy of the high-noise terminals is greater than that of the low-noise terminals, balancing individual transmission rates, as seen in Figs.~\ref{fig:p8_cdf}(top) and \ref{fig:rate_cummean}(top). On the other extent, the algorithm maximizes weighted sum of individual rates for the sumrate utility, where high-noise terminals are worse-off in terms of transmission rates. The algorithm allocates more power to low-noise terminals to increase the total (weighted sum of) throughput, thus we expectedly observe a bias in allocated power towards the low-noise terminals for sumrate utility, as seen in Fig.~\ref{fig:p8_cdf}(bottom).

\vspace{-3pt}
The opportunistic behavior implied above is similar to classical ergodic resource allocation; however, our approach enriches standard ergodic-optimal policies with explicit regulation of power fluctuation risk, which is always feasible and tunable at will, approximating the behavior of a system enforcing a deterministic total power constraint (as in problem \eqref{eqn:problem_formulation_ergodic}) with arbitrary precision.

We now focus on the case of the larger network with $8$ terminals (see Table \ref{tab:sim_parameters2}). As a toy scenario, terminals with distinct noise variances --dashed curves for risk-neutral (RN) scenarios-- exhibit optimal policies affected by respective noise variances for proportional fairness and sumrate utilities, respectively, as shown in Fig.~\ref{fig:p8_cdf}. As also mentioned above, ergodic transmission rates are affected proportionally to the noise variances, as shown in Fig.~\ref{fig:rate_cummean}.

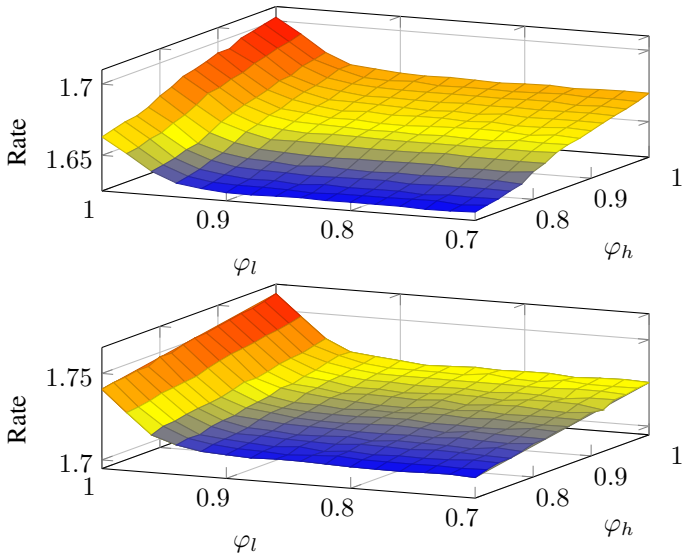
\begin{figure}[t]
\begin{subfigure}[ht]{\linewidth}
\begin{tikzpicture}[trim axis right,baseline]
\begin{axis}[
width=\linewidth,
height=.5\linewidth,
ytick={.8, .9, 1},
ztick={1.65, 1.7},
xlabel={$\varphi_l$},
x dir=reverse,
ylabel={$\varphi_h$},
zlabel={Rate},
xmin=0.7, xmax=1, ymin=.7, ymax=1, zmin=1.625, zmax = 1.71,
xmajorticks=true, ymajorticks=true,
legend style={at={(1,0)},anchor=south east,
nodes={scale=0.4, transform shape}, legend columns = 4}, grid]
 \addplot3[surf] table[x=x, y=y, z=z] {Data-Mesh/mesh_pf_data.txt};
\end{axis}
\end{tikzpicture}
\end{subfigure}
\begin{subfigure}[ht]{\linewidth}
\begin{tikzpicture}[trim axis right,baseline]
\begin{axis}[
width=\linewidth,
height=.5\linewidth,
ytick={.8, .9, 1},
ztick={1.7, 1.75},
xlabel={$\varphi_l$},
x dir=reverse,
ylabel={$\varphi_h$},
zlabel={Rate},
xmin=0.7, xmax=1, ymin=.7, ymax=1, zmin=1.695, zmax = 1.765,
xmajorticks=true, ymajorticks=true,
legend style={at={(1,0)},anchor=south east,
nodes={scale=0.4, transform shape}, legend columns = 4}, grid]
 \addplot3[surf] table[x=x, y=y, z=z] {Data-Mesh/mesh_sr_data.txt};
\end{axis}
\end{tikzpicture}
\end{subfigure}
\caption{Ergodic average rates for the ``realistic" $8$-terminal network, relative to $\varphi_l, \varphi_h$. Top: Proportional Fairness $|$ Bottom: Sumrate.}
\label{fig:surfaces_nominal}
\vspace{12bp}
\end{figure}

\vspace{-4pt}
As a more realistic scenario, we lastly investigate a case where $6$ low-noise and $2$ high-noise terminals are located in the network. Low-noise and high-noise terminals are assigned with common confidence levels respectively, namely $\varphi_l$ and $\varphi_h$. Although the allocated powers are higher for high-noise terminals than low-noise ones for the proportional fairness utility (see Fig.~\ref{fig:p8_cdf}(left)), the difference in ergodic transmission rate is drastically significant, as shown in Fig.~\ref{fig:rate_cummean}(left). We observe otherwise for policies using the sumrate utility, whereas the transmission rate difference is similarly significant. We also investigate the effect of $\varphi_l$ and $\varphi_h$ on the achieved average ergodic rate. Expectedly, the ergodic average rate tops when $\varphi_l$ and $\varphi_h$ are $1$ --corresponding to the risk-neutral ergodic-optimal case-- and reduces slightly as the confidence levels decrease, shown in Fig.~\ref{fig:surfaces_nominal} for proportional fairness and sumrate utilities, respectively. Percentage-wise, the maximal rate optimality loss we have empirically observed is minimal for both utilities, and roughly $4\%$ and $3\%$, respectively. 


\section{Conclusion}
\label{sec:conclusion}
We introduced and investigated a new distributionally robust formulation of a fundamental constrained stochastic resource allocation problem in point-to-point communication networks, strictly regulating the statistical volatility of optimal ergodic power policies. Exploiting the CVaR and within a Lagrangian duality framework, we developed closed-form solutions for all primal variables, and derived stochastic subgradient updates for dual variables. Detailed numerical simulations confirmed the clear effectiveness of the proposed scheme.

\bibliographystyle{IEEEtran}
\bibliography{Files/references}

\begin{thebibliography}{10}
\providecommand{\url}[1]{#1}
\csname url@samestyle\endcsname
\providecommand{\newblock}{\relax}
\providecommand{\bibinfo}[2]{#2}
\providecommand{\BIBentrySTDinterwordspacing}{\spaceskip=0pt\relax}
\providecommand{\BIBentryALTinterwordstretchfactor}{4}
\providecommand{\BIBentryALTinterwordspacing}{\spaceskip=\fontdimen2\font plus
\BIBentryALTinterwordstretchfactor\fontdimen3\font minus \fontdimen4\font\relax}
\providecommand{\BIBforeignlanguage}[2]{{%
\expandafter\ifx\csname l@#1\endcsname\relax
\typeout{** WARNING: IEEEtran.bst: No hyphenation pattern has been}%
\typeout{** loaded for the language `#1'. Using the pattern for}%
\typeout{** the default language instead.}%
\else
\language=\csname l@#1\endcsname
\fi
#2}}
\providecommand{\BIBdecl}{\relax}
\BIBdecl

\bibitem{ribeiro_optimal_2012}
A.~Ribeiro, ``Optimal resource allocation in wireless communication and networking,'' \emph{EURASIP Journal on Wireless Commun. and Networking}, vol. 2012, no.~1, Aug. 2012, publisher: Springer Science and Business Media LLC.

\bibitem{li_efficient_2021}
Y.~Li, D.~Guo, Y.~Zhao, X.~Cao, and H.~Chen, ``Efficient {Risk}-{Averse} {Request} {Allocation} for {Multi}-{Access} {Edge} {Computing},'' \emph{IEEE Communications Letters}, vol.~25, no.~2, pp. 533--537, 2021.

\bibitem{bennis_ultrareliable_2018}
M.~Bennis, M.~Debbah, and H.~V. Poor, ``Ultrareliable and {Low}-{Latency} {Wireless} {Communication}: {Tail}, {Risk}, and {Scale},'' \emph{Proceedings of the IEEE}, vol. 106, no.~10, pp. 1834--1853, 2018.

\bibitem{vu_ultra_reliable_2018}
T.~K. Vu, M.~Bennis, M.~Debbah, M.~Latva-aho, and C.~S. Hong, ``Ultra-{Reliable} {Communication} in {5G} {mmWave} {Networks}: {A} {Risk}-{Sensitive} {Approach},'' \emph{IEEE Commun. Letters}, vol.~22, no.~4, pp. 708--711, 2018.

\bibitem{yang_localization_assisted_2024}
Z.~Yang, S.~Yang, H.~Zhang, B.~Di, X.~Li, X.~Hou, and L.~Song, ``Localization-assisted {Communication} for {RIS}-aided {Distributed} {MIMO} {Systems} {Using} {Distributionally} {Robust} {Optimization},'' in \emph{2024 {IEEE}/{CIC} {International} {Conference} on {Communications} in {China} ({ICCC})}, 2024, pp. 797--802.

\bibitem{batewela_risk_sensitive_2020}
S.~Batewela, C.-F. Liu, M.~Bennis, H.~A. Suraweera, and C.~S. Hong, ``Risk-{Sensitive} {Task} {Fetching} and {Offloading} for {Vehicular} {Edge} {Computing},'' \emph{IEEE Communications Letters}, vol.~24, no.~3, pp. 617--621, 2020.

\bibitem{azari_risk-aware_2019}
A.~Azari, M.~Ozger, and C.~Cavdar, ``Risk-{Aware} {Resource} {Allocation} for {URLLC}: {Challenges} and {Strategies} with {Machine} {Learning},'' \emph{IEEE Communications Magazine}, vol.~57, no.~3, pp. 42--48, 2019.

\bibitem{chai_u_reliable_v2x_2024}
G.~Chai, W.~Wu, Q.~Yang, M.~Qin, Y.~Wu, and F.~R. Yu, ``{Platoon Partition and Resource Allocation for Ultra-Reliable V2X Networks},'' \emph{IEEE Transactions on Vehicular Technology}, vol.~73, no.~1, pp. 147--161, 2024.

\bibitem{wang_jspa_2023}
P.~Wang, W.~Wu, J.~Liu, G.~Chai, and L.~Feng, ``{Joint Spectrum and Power Allocation for V2X Communications With Imperfect CSI},'' \emph{IEEE Transactions on Vehicular Technology}, vol.~72, no.~12, pp. 16\,338--16\,353, 2023.

\bibitem{wu_rra_v2x_2023}
W.~Wu, P.~Wang, Y.~Fan, J.~Liu, R.~Liu, and W.~Xia, ``{Robust Resource Allocation for RIS-aided V2X Communications with Imperfect CSI},'' in \emph{2023 IEEE 98th Vehicular Technology Conference (VTC2023-Fall)}, 2023, pp. 1--6.

\bibitem{cui_dro_mec_2023}
C.~Cui, Z.~Jia, C.~Dong, Z.~Ling, J.~You, and Q.~Wu, ``{Distributionally Robust Chance-Constrained Optimization for Hierarchical UAV-based MEC},'' in \emph{IEEE INFOCOM 2023 - IEEE Conference on Computer Communications Workshops (INFOCOM WKSHPS)}, 2023, pp. 1--6.

\bibitem{wu_rra_vc_2021}
W.~Wu, R.~Liu, Q.~Yang, and T.~Q.~S. Quek, ``{Robust Resource Allocation for Vehicular Communications With Imperfect CSI},'' \emph{IEEE Transactions on Wireless Communications}, vol.~20, no.~9, pp. 5883--5897, 2021.

\bibitem{ling_dro_offloading_2020}
Z.~Ling, F.~Hu, Y.~Zhang, F.~Gao, and Z.~Han, ``{Distributionally Robust Chance-Constrained Optimization for Communication and Offloading in WBANs},'' in \emph{GLOBECOM 2020 - 2020 IEEE Global Communications Conference}, 2020, pp. 1--6.

\bibitem{lie_dro_jammer_2019}
X.~Liu, Y.~Gao, L.~Wang, N.~Sha, and S.~Wang, ``{Distributionally Robust Optimization for Secure Transmission With Assisting Jammer in MISO Downlink Networks},'' \emph{IEEE Communications Letters}, vol.~23, no.~2, pp. 338--341, 2019.

\bibitem{gong_dro_relay_2016}
S.~Gong, S.~X. Wu, A.~M.-C. So, and X.~Huang, ``{Distributionally Robust Relay Beamforming in Wireless Communications},'' in \emph{Proceedings of the 19th ACM International Conference on Modeling, Analysis and Simulation of Wireless and Mobile Systems}.\hskip 1em plus 0.5em minus 0.4em\relax Association for Computing Machinery, 2016, p. 254–261.

\bibitem{li_dro_beamforming_2014}
Q.~Li, A.~M.-C. So, and W.-K. Ma, ``{Distributionally robust chance-constrained transmit beamforming for multiuser MISO downlink},'' in \emph{2014 IEEE International Conference on Acoustics, Speech and Signal Processing (ICASSP)}, 2014, pp. 3479--3483.

\bibitem{li_ra_dro_mec_2023}
Z.~Li and P.~Chen, ``{Risk-Aware Distributionally Robust Optimization for Mobile Edge Computation Task Offloading in the Space–Air–Ground Integrated Network},'' \emph{Sensors}, vol.~23, no.~12, 2023.

\bibitem{yaylali_robust_2023}
G.~Yaylali and D.~Kalogerias, ``Robust and {Reliable} {Stochastic} {Resource} {Allocation} via {Tail} {Waterfilling},'' in \emph{2023 {IEEE} 24th {International} {Workshop} on {Signal} {Processing} {Advances} in {Wireless} {Communications} ({SPAWC})}, 2023, pp. 256--260.

\bibitem{kalogerias_strongdual_2023}
D.~Kalogerias and S.~Pougkakiotis, ``{Strong Duality in Risk-Constrained Nonconvex Functional Programming},'' \emph{arXiv preprint arXiv:2206.11948}, 2023.

\bibitem{yaylali_stochastic_2024}
G.~Yaylali and D.~Kalogerias, ``Stochastic {Resource} {Allocation} via {Dual} {Tail} {Waterfilling},'' in \emph{2024 58th {Annual} {Conference} on {Information} {Sciences} and {Systems} ({CISS})}, Mar. 2024, pp. 1--6.

\bibitem{rahimian_dro_2022}
H.~Rahimian and S.~Mehrotra, ``{Frameworks and Results in Distributionally Robust Optimization},'' \emph{Open Journal of Mathematical Optimization}, vol.~3, p. 1–85, Jul. 2022.

\bibitem{lin_dro_2022}
F.~Lin, X.~Fang, and Z.~Gao, ``{Distributionally Robust Optimization: A review on theory and applications},'' \emph{{Numerical Algebra, Control \& Optimization}}, vol.~12, no.~1, p. 159, Mar. 2022.

\bibitem{shapiro_lectures_2021}
A.~Shapiro, D.~Dentcheva, and A.~Ruszczynski, \emph{Lectures on {Stochastic} {Programming}: {Modeling} and {Theory}}, 3rd~ed.\hskip 1em plus 0.5em minus 0.4em\relax Society for Industrial and Applied Mathematics, Jul. 2021.

\bibitem{kuhn_dro_2024}
D.~Kuhn, S.~Shafiee, and W.~Wiesemann, ``{Distributionally Robust Optimization},'' \emph{arXiv preprint arXiv:2411.02549}, 2024.

\bibitem{ang_dual_2017}
M.~Ang, J.~Sun, and Q.~Yao, ``On the dual representation of coherent risk measures,'' \emph{{Annals of Operations Research}}, vol. 262, no.~1, p. 29–46, Feb. 2017.

\bibitem{rockafellar_optimization_2000}
R.~T. Rockafellar and S.~Uryasev, ``Optimization of conditional value-at-risk,'' \emph{The Journal of Risk}, vol.~2, no.~3, pp. 21--41, 2000, publisher: Infopro Digital Services Limited.

\bibitem{ribeiro_ergodic_2010}
A.~Ribeiro, ``Ergodic {Stochastic} {Optimization} {Algorithms} for {Wireless} {Communication} and {Networking},'' \emph{IEEE Trans. on Signal Processing}, vol.~58, no.~12, pp. 6369--6386, 2010.

\bibitem{ribeiro_separation_2010}
A.~Ribeiro and G.~B. Giannakis, ``Separation {Principles} in {Wireless} {Networking},'' \emph{IEEE Trans. on Information Theory}, vol.~56, no.~9, pp. 4488--4505, 2010.

\bibitem{cover_elements_2005}
T.~M. Cover and J.~A. Thomas, \emph{Elements of {Information} {Theory}}.\hskip 1em plus 0.5em minus 0.4em\relax Wiley, Apr. 2005.

\end{thebibliography}

\end{document}